\documentclass[journal=accacs,manuscript=article,layout=twocolumn]{achemso}
\usepackage[version=3]{mhchem} 

\newcommand\tabcaption{\def\@captype{table}\caption}
\newcommand\figcaption{\def\@captype{figure}\caption}
\usepackage{amsmath}
\usepackage{chemformula}    
\usepackage{bm}
\usepackage{multirow}
\usepackage{mdframed}
\usepackage{rotating}
\usepackage{booktabs}
\usepackage{etoolbox}
\usepackage{enumitem}
\usepackage{subcaption}
\usepackage[nonumberlist]{glossaries}
\mciteErrorOnUnknownfalse
\usepackage{xcolor}
\usepackage{comment}
\usepackage{array}
\usepackage[hidelinks]{hyperref}
\newcounter{magicrownumbers}
\preto\tabular{\setcounter{magicrownumbers}{0}}
\usepackage{todonotes}
\usepackage{siunitx}

\newcommand{\mr}{\multirow}

\newcommand{\mc}{\multicolumn}

\newcommand{\dataset}[1]{OC22}
\newcommand{\ocpurl}[0]{\url{http://opencatalystproject.org}}
\newcommand{\modelurl}[0]{\url{https://github.com/Open-Catalyst-Project/ocp/blob/main/MODELS.md}}
\newcommand{\baselinesurl}[0]{\url{https://github.com/Open-Catalyst-Project/ocp}}
\newcommand{\dataseturl}[0]{\url{https://github.com/Open-Catalyst-Project/Open-Catalyst-Dataset}}

\newacronym{DFT}{DFT}{Density Functional Theory}
\newacronym{SI}{SI}{Supporting Information}
\newacronym{PES}{PES}{Potential Energy Surface}
\newacronym{QE}{QE}{Quantum Espresso}
\newacronym{VASP}{VASP}{Vienna Ab initio Simulation Package}
\newacronym{S2EF}{\textit{S2EF}}{Structure to Energy and Forces}
\newacronym{RS2EF}{\textit{S2EF-Total}}{Structure to Total Energy and Forces}
\newacronym{RIS2RE}{\textit{IS2RE-Total}}{Initial Structure to Total Relaxed Energy}
\newacronym{IS2RE}{\textit{IS2RE}}{Initial Structure to Relaxed Energy}
\newacronym{IS2RS}{\textit{IS2RS}}{Initial Structure to Relaxed Structure}
\newacronym{ID}{ID}{In-Domain}
\newacronym{OOD}{OOD}{Out-of-Domain}
\newacronym{OC20}{OC20}{Open Catalyst 2020 Dataset}
\newacronym{ML}{ML}{machine learning}
\newacronym{GNNs}{GNNs}{Graph Neural Networks}
\newacronym{GNN}{GNN}{Graph Neural Network}
\newacronym{MD}{MD}{\textit{ab initio} Molecular Dynamics}
\newacronym{OER}{OER}{Oxygen Evolution Reaction}
\newacronym{MAE}{MAE}{Mean Absolute Error}

\newacronym{EFwT}{EFwT}{Energy and Forces within Threshold}
\newacronym{EwT}{EwT}{Energy within Threshold}
\newacronym{ADwT}{ADwT}{Average Distance within Threshold}
\newacronym{FbT}{FbT}{Force below Threshold}
\newacronym{AFbT}{AFbT}{Average Force below Threshold}
\newacronym{OC22}{OC22}{Open Catalyst 2022}
\newacronym{MvK}{MvK}{Mars-van Krevelen}
\glsaddall

\newcommand{\fair}{Fundamental AI Research, Meta AI}
\newcommand{\coa}{Indicates equal contributions}
\newcommand{\cmu}{Department of Chemical Engineering, Carnegie Mellon University}
\newcommand{\scott}{Scott Institute for Energy Innovation, Carnegie Mellon University}
\newcommand{\uoftdpes}{Department of Physical and Environmental Science, University of Toronto}
\newcommand{\uoftdece}{Department of Electrical and Computer Engineering, University of Toronto}
\newcommand{\uoftmse}{Department of Materials Science and Engineering, University of Toronto}

\author{Richard Tran}
\affiliation{\coa}
\alsoaffiliation{\cmu}
\author{Janice Lan}
\affiliation{\coa}
\alsoaffiliation{\fair}
\author{Muhammed Shuaibi}
\affiliation{\coa}
\alsoaffiliation{\cmu}
\alsoaffiliation{\fair}
\author{Brandon M. Wood}
\affiliation{\coa}
\alsoaffiliation{\fair}
\author{Siddharth Goyal}
\affiliation{\coa}
\alsoaffiliation{\fair}
\author{Abhishek Das}
\affiliation{\fair}
\author{Javier Heras-Domingo}
\affiliation{\cmu}
\author{Adeesh Kolluru}
\affiliation{\cmu}
\author{Ammar Rizvi}
\affiliation{\fair}
\author{Nima Shoghi}
\affiliation{\fair}
\author{Anuroop Sriram}
\affiliation{\fair}
\author{F\'elix Therrien}
\affiliation{\uoftdece}
\alsoaffiliation{\uoftdpes}
\author{Jehad Abed}
\affiliation{\uoftdece}
\alsoaffiliation{\uoftmse}
\author{Oleksandr Voznyy}
\affiliation{\uoftdpes}
\author{Edward H. Sargent}
\affiliation{\uoftdece}
\author{Zachary Ulissi}
\affiliation{\scott}
\alsoaffiliation{\cmu}
\email{zulissi@andrew.cmu.edu}
\author{C.~Lawrence Zitnick}
\affiliation{\fair}
\email{zitnick@meta.com}

\title[]
  {The Open Catalyst 2022 (OC22) Dataset  and Challenges for Oxide Electrocatalysts}

\abbreviations{IR,NMR,UV}

\let\oldmaketitle\maketitle
\let\maketitle\relax

\begin{document}



\twocolumn[
\begin{@twocolumnfalse}
\oldmaketitle
\begin{abstract}
\noindent The development of machine learning models for electrocatalysts requires a broad set of training data to enable their use across a wide variety of materials. One class of materials that currently lacks sufficient training data is oxides, which are critical for the development of \gls{OER} catalysts. To address this, we developed the \gls{OC22} dataset, consisting of 62,331 \gls{DFT} relaxations ($\sim$9,854,504 single point calculations) across a range of oxide materials, coverages, and adsorbates. We define generalized total energy tasks that enable property prediction beyond adsorption energies; we test baseline performance of several graph neural networks; and we provide pre-defined dataset splits to establish clear benchmarks for future efforts. In the most general task, GemNet-OC sees a $\sim$36\% improvement in energy predictions when combining the chemically dissimilar \gls{OC20} and \gls{OC22} datasets via fine-tuning. Similarly, we achieved a ${\sim}19\%$ improvement in total energy predictions on \gls{OC20} and a $\sim$9\% improvement in force predictions in \gls{OC22} when using joint training. We demonstrate the practical utility of a top performing model by capturing literature adsorption energies and important \gls{OER} scaling relationships. We expect \gls{OC22} to provide an important benchmark for models seeking to incorporate intricate long-range electrostatic and magnetic interactions in oxide surfaces. Dataset and baseline models are open sourced, and a public leaderboard is available to encourage continued community developments on the total energy tasks and data.
\end{abstract}

\end{@twocolumnfalse}
]

\clearpage
\section{Introduction}
Advances are needed in technologies to produce, store, and use low-carbon-intensity energy. Renewable energy is often produced by intermittent sources (e.g. sunlight, wind, or tides) so efficient grid-scale storage is required to transfer power from times of excess generation to times of excess demand. There are a number of promising storage techniques including the conversion of renewable energy to a chemical form, e.g. water splitting to \ch{H2}, or \ch{CO2} conversion to liquid fuels and high-value chemical feedstock. Inorganic oxides are abundant electrocatalysts that are extensively used in these applications. However, the complex nature of oxide surfaces compared to simpler metals present a number of challenges to catalyst design. Developing generalizable machine learning methods to quickly and accurately predict the activity and stability of oxide catalysts would have a major impact on renewable energy storage and utilization.

As a motivating example of the need and challenges for oxide electrocatalysts, we consider water splitting for the generation of clean H$_2$; an energy-dense fuel that is used in fuel cells or ammonia synthesis. Electrochemical water splitting consists of two coupled half-reactions, 
\begin{align*}
    \centering
        \text{OER:}&& \ce{2 H2O &-> O2 + 4(H^{+} + e^{-})}\\
        \text{HER:}&& \ce{4(H^{+} + e^{-}) &-> 2H2}\\
        \hline
        &&\ce{2H2O &-> O2 + H2}, 
\end{align*}
which split two water molecules to evolve H$_2$ and O$_2$ gas. This process is extremely energy intensive. The \gls{OER} overpotential is the larger contributor to the inefficiency of this reaction; it is quite complicated due to bond rearrangements and the formation of an \ce{O-O} bond. Water splitting typically uses very harsh acidic conditions to reduce gas solubility and improve proton conductivity, and for which high performance proton exchange membranes are widely available. Unfortunately, for these conditions there are very few known materials that are stable and active, except extremely expensive metal oxides, such as those using Ir\cite{Trasatti1980}. Currently, there are significant efforts to design complex multi-component oxide \gls{OER} catalysts to reduce the cost and improve their activity and stability \cite{Jamesh2018, Yuan2020}. Computational chemistry can play a critical role in helping screen, discover, and understand such materials. 

Computational methods can be used to predict the activity and stability of a proposed oxide catalyst, but these techniques are significantly more complicated than for metal catalysts and present many additional challenges. First, there are many oxide polymorphs (crystal structures) for any given chemical composition that must be considered to identify the most stable catalyst structure\cite{flores2020active}. Second, the surface of an oxide catalyst is often prone to reconstruction, leaching, doping, and defects \cite{doi:10.1021/acs.chemrev.0c01060}. Third, the environment can lead to a number of possible surface terminations. Fourth, it is difficult to determine a catalyst's active site and there are often multiple competing mechanisms to consider \cite{Gonzalez2021}. To add to these challenges, computational chemistry methods such as the widely-used Generalized Gradient Approximation (GGA) are less accurate for oxide materials due to the strong electron correlation and complicated electronic structure. Large system sizes and the likelihood of long-range electrostatic or magnetic interactions also result in slower convergence. These additional configurational and computational complexities make the creation of datasets and machine learning models for oxides significantly more expensive and challenging, leading to much fewer and smaller datasets  than for metal systems (see \cite{andersen2021adsorption} for a sample of representative datasets in catalysis). 

\begin{figure*}[t]
    \centering
    \includegraphics[width=0.8\textwidth]{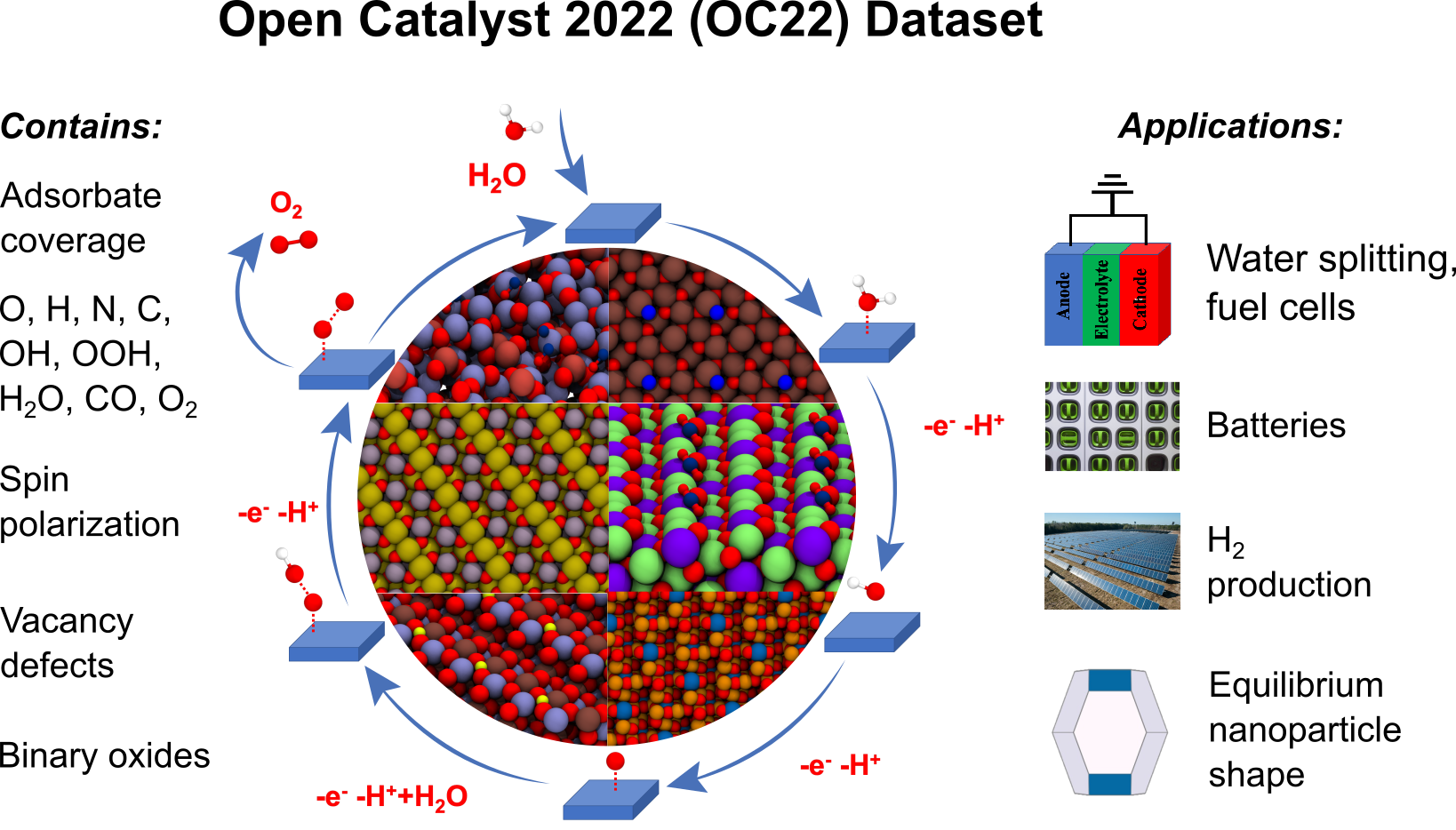}
    \caption{Overview of the contents and impact areas of the \gls{OC22} dataset. The water nucleophilic attack mechanism is highlighted for the \gls{OER} reaction, with \ch{H2O} and \ch{O2} as reactants and products, respectively. Inset images are a random sample of the dataset.}
    \label{fig:oc22_overview}
\end{figure*}

We address these challenges by generating a large oxide dataset to accelerate the development of \gls{ML} models for materials design and discovery.  The training of accurate and generalizable ML models requires large datasets. For example, the \gls{OC20} dataset \cite{chanussot2021open} (ca. 250 million single-point calculations) considered different adsorbates (small adsorbates, C1/C2 compounds, and N/O-containing intermediates) on top of randomly sampled low Miller index facets of stable materials from the Materials Project\cite{Jain2013}, but did not include metal oxide materials due to the complexities above. The release of the \gls{OC20} dataset helped enable rapid advances in the accuracy and generalizability of \gls{GNN} models \cite{oc20_perspective}, with decreases of 55+\% in the key S2EF metrics in the first two years. Initial baseline models like CGCNN\cite{xie2018crystal} and SchNet\cite{schutt2017schnet} focused on local environment representations. Key advances since then include invariant angular interactions (DimeNet/DimeNet++ \cite{klicpera2020directional, klicpera2020fast}), faster and more accurate but non-energy conserving models (ForceNet\cite{hu2021forcenet} and SpinConv\cite{shuaibi2021rotation}), and triple/quadruplet interactions (GemNet-dT\cite{gasteiger2021gemnet}, GemNet-XL\cite{sriram2022towards}, and GemNet-OC\cite{gasteiger2022graph}). Other approaches include the use of transformers (3D-Graphormer\cite{ying2021transformers}) and more effective augmentation and learning strategies (Noisy-Nodes\cite{godwin2021simple}). These and further advances are necessary to accurately predict properties of complex structures such as oxide systems.

In this work, we present the Open Catalyst 2022 (OC22) dataset (Figure \ref{fig:oc22_overview}) for the oxygen evolution reaction and oxide electrocatalysts more generally, as well as accompanying tasks and \gls{GNN} baseline models. \gls{OC22} is intended to complement \gls{OC20}, which did not contain any oxide materials, and further enable the development of generalizable \gls{ML} models for catalysis. This dataset spans the configurational complexity for oxide surfaces described above, including varying surface terminations, adsorbate+slab configurations and coverage, and non-stoichiometric substitutions and vacancies. To encompass the additional complexities in this dataset, we also expand on the primary tasks in \gls{OC20} to include the \gls{DFT} total energy as a target. A more general property, \gls{DFT} total energy offers the ability to address potential applications beyond those that just require simple adsorption energies.

With the creation of new datasets, the question arises of whether the data in them is complementary to other datasets for training \gls{ML} models (see recent reviews for a perspective of catalysts informatics\cite{Takahashi2019, medford2018extracting, schlexer2019machine}). This is especially important when consolidating data with a variety of computational methods in anthological dataset collections such as the Catalysis-Hub\cite{Winther2019}, Catalyst Acquisition by Data Science\cite{Fujima2020}, and the NFDI4Cat consortium \cite{Wulf2021}. For instance, models can be trained jointly using multiple datasets, or transfer learning may be used to train a model on a larger dataset and fine-tuned on a smaller dataset. Recently, the \gls{OC20} dataset enabled the catalysis community to use transfer learning to improve model performance \cite{kolluru2022transfer} on other smaller datasets. The small molecules and drug discovery communities have seen success in using transfer learning to transfer between varying levels of electronic structure calculations  \cite{smith2019approaching} or between related tasks\cite{dai2009eigentransfer, rosenbaum2013inferring, turki2017transfer}. In this work, we explore the extent \gls{OC20} can aid \gls{OC22} via transfer learning or by jointly training on both datasets. 

We train a variety of leading \gls{GNN} models on two related proposed community challenges for \gls{OC22}: (1) predict the \gls{DFT} total energy and force for a given structure and (2) predict the \gls{DFT} relaxed total energy given an initial structure. We also evaluate our models' performance on the established task of predicting the relaxed structure given an initial structure. The dataset is split into train/validation/test splits indicative of the situation commonly found in catalysis where the properties of unseen crystal compositions need to be predicted. Splits contain a combination of isolated surfaces (a.k.a slab) and surface with adsorbate (a.k.a adsorbate+slab)  systems. All baseline models, data loaders and training scripts for each of these tasks are available at \baselinesurl{}. While we focus on a subset of tasks, models capable of solving these tasks on the \gls{OC22} dataset will likely be able to address numerous related catalysis problems.
\section{The \gls{OC22} Dataset}

\gls{OC22} is designed to provide a training dataset for constructing generalized models to aid in predicting catalytic reactions on oxide surfaces. To achieve this, we built the dataset in four stages: (1) bulk selection, (2) surface selection, (3) initial structure generation, and (4) structure relaxation. The dataset contains slabs and adsorbate+slabs, 19,142 and 43,189 systems, respectively. This resulted in 9,854,504 single-point calculations, each of which yielded forces and energies that were later partitioned into suitable train, validation, and test validation splits. We prioritized diversity in composition, surface termination, and adsorbate configurations in constructing our dataset to ensure that our models can generalize well.  As a result of our emphasis on creating an unbiased and diverse dataset, \gls{OC22} structures may not always be the most stable or pertinent for a particular reaction pathway of interest - data still meaningful for building generalizable models.  All source code used to generate the adsorbate configurations will be provided in the Open Catalyst Dataset repository at \dataseturl.

\subsection{Bulk selection}

We begin by confining our set of bulk oxide materials to 4,728 unary (\ce{A_xO_y}) and binary (\ce{A_xB_yO_z}) metal-oxides from the Materials Project\cite{Jain2013} where A and B are metals. These oxides can be composed of any combination of metals or semi-metals listed in the \gls{SI}. In our list of 51 metals, \ch{Ce} and \ch{Lu} were the only lanthanides considered due to the utility of cerium-based oxide compounds in catalytic reactions\cite{Song2021, Dey2020} and to add additional variety with lutetium-based oxides. For each chemical system, we considered bulk materials with the top five lowest energies above hull with less than 150 atoms to provide equal chemical distribution and diversity in our set of oxides. We note that under this criteria, some materials may exhibit an energy above hull exceeding 0.1 eV/atom (the threshold initially used in OC20). In addition to chemical diversity, we also included materials with a variety of electronic band gaps ($E_G$). Table~\ref{tab:comp_sample_table} lists the number of metallic ($E_G = 0 ~\text{eV}$), semiconducting ($0~\text{eV} < E_G < 3.2~\text{eV}$), and insulating ($E_G > 3.2~\text{eV}$) materials considered in our dataset (all electronic properties were derived from the Materials Project). Many oxides such as \ce{TiO2} are also useful for photocatalysis which typically requires semiconducting properties to allow for photoelectron excitation.  We also considered 173 unary and binary rutile structures.

Our selection criteria for bulk oxides prioritized chemical diversity over stability. We acknowledge that many of the materials we selected are not electrochemically stable which is a prerequisite for viable electrocatalytic materials. Pourbaix analysis have previously demonstrated that only oxides composed of 26 of the 51 elements we considered are relatively stable under aqueous conditions\cite{wang2020acid}.

We also ignored the fact that certain chemical systems have a far greater set of distinct bulk structures than others. For instance, the Materials Project database has reported over 300 entries for chemical systems such as Ti-O and Mn-Li-O while no entries were reported for 200 chemical systems (see the \gls{SI}). Other databases such as the Automatic-Flow\cite{Curtarolo2012} and Open Quantum Materials Database\cite{saal2013materials} have also made significant efforts in exploring oxides and contain chemical systems unexplored in the Materials Project. However, to ensure all oxides were obtained using a consistent methodology and open source licensing, we extracted entries from the Materials Project only. 

\subsection{Surface selection}

\begin{figure}[h]
    \centering
    \includegraphics[width=\columnwidth]{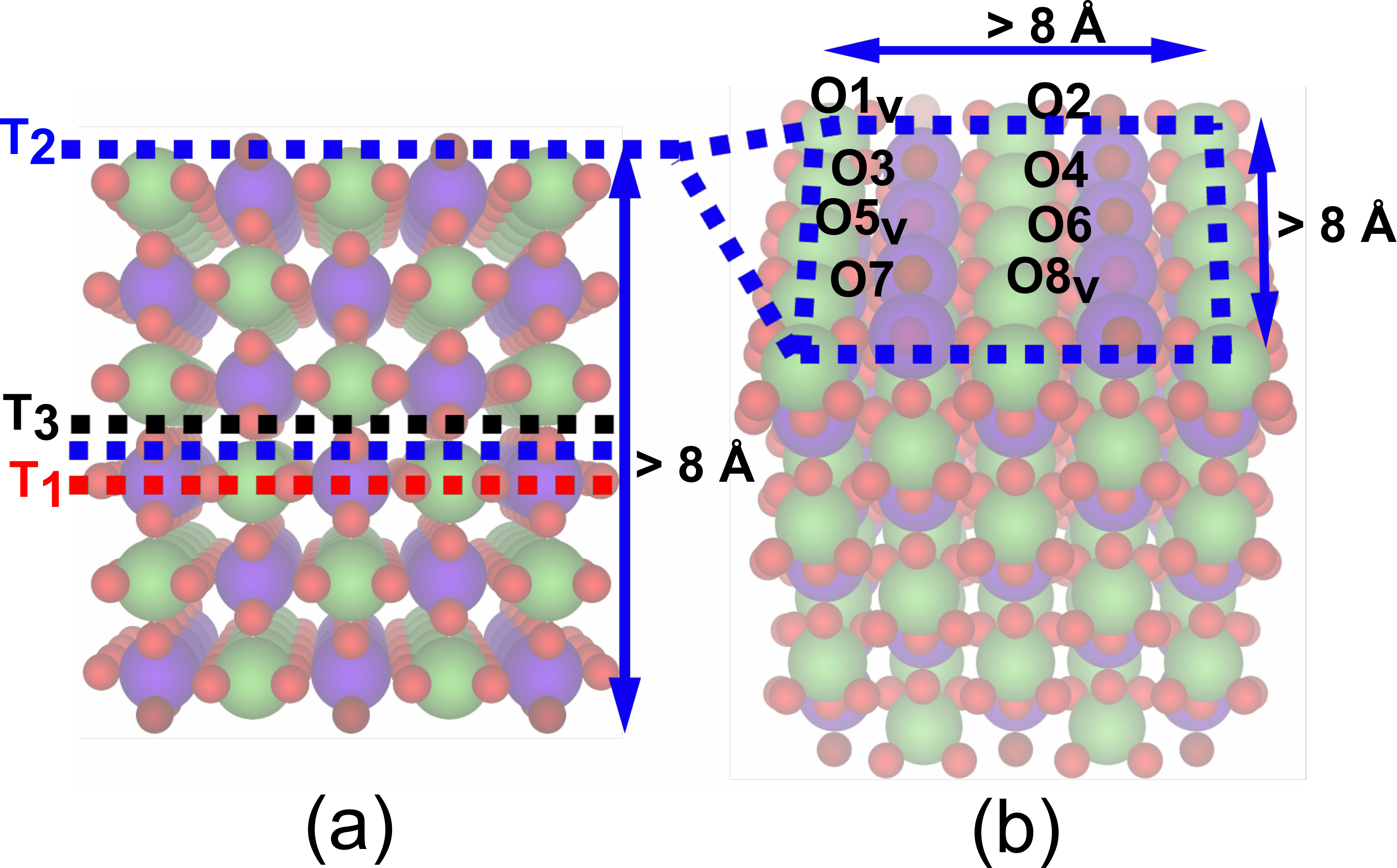}
    \caption{Construction of rutile (110) slabs and adsorbate+slabs. (a) Dashed lines indicate the different possible terminations ($T_{1}, T_{2}$ and $T_{3}$). The slab is symmetric about $T_{3}$. (b) The $T_{2}$ terminated surface with its periodic boundary (blue dashed lines) contains 8 oxygen sites. Random removal of 3 surface oxygen (dark red) creates vacancy defects (transparent).
}
    \label{fig:slab_construction}
\end{figure}

We constructed our dataset by first randomly sampling 4,286 bulk oxides from our original bulk oxide set of 4,728. We limited our dataset to slabs of less than 250 atoms. We construct each slab and adsorbate+slab using the process shown in Figure~\ref{fig:slab_construction}. Given a random oxide selected from our bulk dataset, we enumerate through all possible surface terminations with a maximum Miller index less than or equal to 3. As with Figure~\ref{fig:slab_construction}(a) all slabs are capped with the same terminating surface regardless of stoichiometry. We randomly select one termination which we replicated to a depth of at least 8 Å and a width in each cross-sectional direction of at least 8 Å.

Next we decorated the surface of the slab with a random number of oxygen vacancies which can act as active sites for reactions such as \ce{CO2} capture\cite{Liu2018a} and OER\cite{Asnavandi2018, Lopes2021}. To do so, we first identify all existing oxygen lattice sites on the surface as with Figure~\ref{fig:slab_construction}(b).
We then select a random number of surface oxygen to remove ranging from 0 (no vacancies) to all surface oxygen. We do the same on the other surface to maintain surface symmetry and avoid the manifestation of non-physical dipole moments which can lead to diverging DFT energies.

The \gls{SI} provides the chemical space distribution of all slabs and adsorbate+slabs successfully calculated in the dataset. Table~\ref{tab:comp_sample_table} summarizes the distribution of elemental composition, crystal structures, bulk band gap, and number of components of the entire dataset of slabs and adsorbate+slabs.

\begin{table}[ht!]
\caption{Overview of the chemical, structural and adsorbate composition of the entire dataset of slabs and adsorbate+slabs.}
\label{tab:comp_sample_table}
\begin{tabular}{rc}
\toprule
  \multicolumn{2}{c}{\textbf{Chemical formula}} \\
  Unary (\ce{A_xO_y}) & 6,190 \\
  Binary (\ce{A_xB_yO_z}) & 56,141 \\
\midrule
  \multicolumn{2}{c}{\textbf{Elements sampled}} \\
  Alkali & 13,541 \\
  Alkaline & 13, 974 \\
  p-block metals & 14,029 \\
  Metalloids &  8,292 \\
  Transition metals & 48,561 \\
\midrule

  \multicolumn{2}{c}{\textbf{Crystal structures}} \\
  Triclinic & 6,214 \\
  Monoclinic & 16,294 \\
  Orthorhombic & 7,258 \\
  Tetragonal (Rutile) & 11,550 (4,318)\\
  Trigonal & 4,411 \\
  Hexagonal & 2,680 \\
  Cubic & 9,606 \\
\midrule

  \multicolumn{2}{c}{\textbf{Band gaps}} \\
  $E_G = 0 ~\text{eV}$ & 1,366 \\
  $ 0 ~\text{eV} < E_G < 3.2 ~\text{eV}$ & 2,591 \\
  $E_G > 3.2 ~\text{eV}$ & 598 \\
\midrule

  \multicolumn{2}{c}{\textbf{Adsorbates}} \\
  O & 10,816 \\
  H & 5,298 \\
  N & 4,000 \\
  C & 3,905 \\
  OH & 4,092 \\
  OOH & 4,424 \\
  \ce{H2O} & 4,846 \\
  CO & 3,994 \\
  \ce{O2} & 1,814 \\
\midrule

 \multicolumn{2}{c}{\textbf{Calc. with PBE+U}: 20,812} \\
\bottomrule
\end{tabular}
\end{table}

\subsection{Initial Structure Generation}
\begin{figure*}
    \centering
    \includegraphics[width=\textwidth]{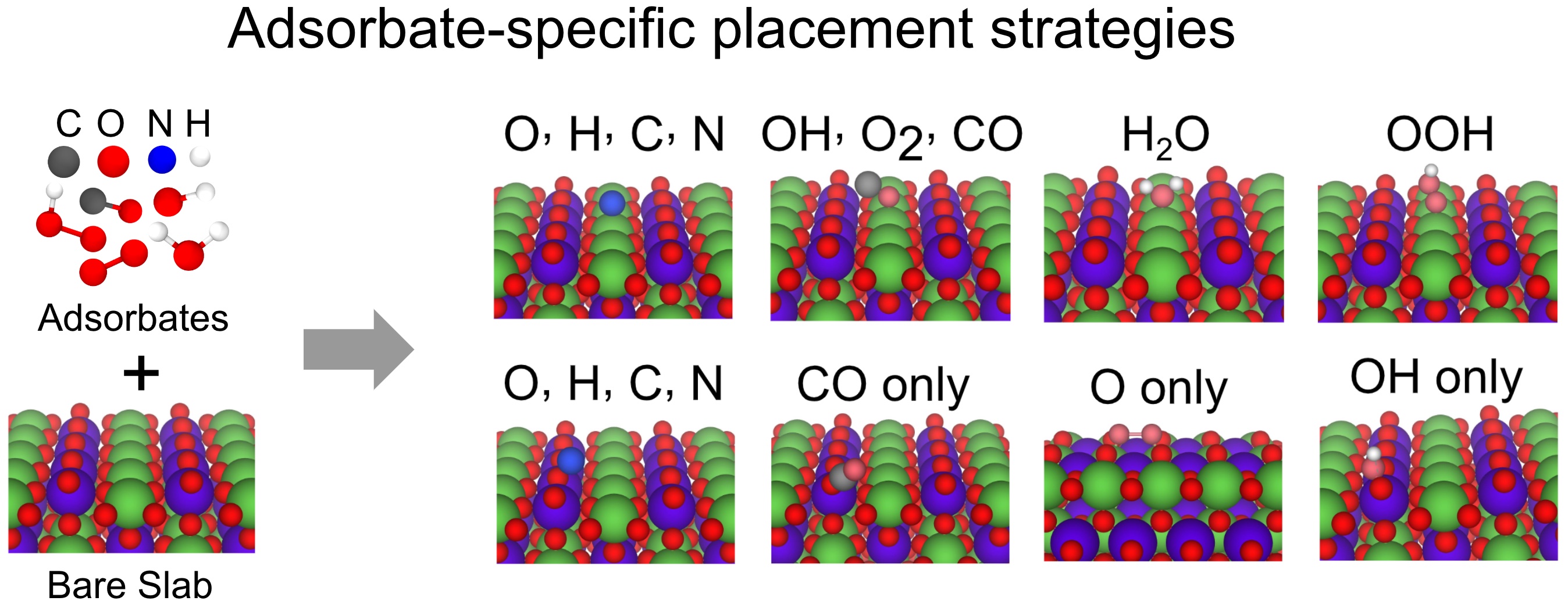}
    \caption{Overview of the adsorbate specific placement strategies. Adsorbates include \ch{H}*, \ch{O}*, \ch{N}*, \ch{C}*, \ch{OOH}*, \ch{OH}*, \ch{OH2}*, \ch{O2}*, and \ch{CO}* (left). Adsorbates can either bind to undercoordinated surface metals (first row of strategies) or to surface oxygen to form new intermediates (second row).}
    \label{fig:adsorbates}
\end{figure*}

To construct our adsorbate+slab, we first randomly sample one adsorbate from the set shown in Figure~\ref{fig:adsorbates}. This adsorbate set includes \ce{O}*, \ce{OH}*, \ce{OH2}*, \ce{OOH}*, and \ce{O2}* which are the intermediates in the proposed reaction mechanisms of OER. To expand the possible chemistry of adsorbates on oxides beyond OER, we also included monatomic \ch{H}*, \ch{O}*, \ch{N}*, and \ch{C}*, as well as \ce{CO}*. Table~\ref{tab:comp_sample_table} shows the distribution of the 9 sampled adsorbates across the dataset.

We then determine the coverage of our random adsorbate on our randomly constructed slab. In contrast to the \gls{OC20} dataset, here we allow for more than one adsorbate of the same type to bind to the surface. The adsorbate can bind to three types of sites: the surface oxygen, the under-coordinated surface metal, or an oxygen vacancy. The maximum number of adsorbates allowed on the surface is limited by the sum of these three types of sites. However we also ensure that all adsorbates are always separated by a distance greater than the M-O bond of the host material to avoid adsorbate overcrowding. 

In this effort, we implemented specific strategies for placing adsorbates on the aforementioned surface sites as shown in Figure~\ref{fig:adsorbates}. The first row of placement strategies demonstrates that all adsorbates are able to bind to any undercoordinated surface metal at the lattice position of oxygen. This includes lattice positions of vacancies introduced during slab generation. An adsorbate containing oxygen will always bind to the metal via the oxygen atom as shown for \ce{OH}*, \ce{O2}*, \ce{CO}*, \ce{H2O}* and OOH*. We also considered intermediates that arise due to formation of oxygen dimers which play a role in one of the possible mechanisms of OER\cite{Gonzalez2021, Dau2010}. In this configuration, a pair of monatomic oxygen atoms can adsorb on to adjacent undercoordinated metals to form a dimer of 1.68~\AA which is longer than the bond length of \ce{O2}*.

The second row demonstrate how specific molecules that are able to form new molecules with the addition of oxygen can also bind to existing surface oxygen. For example, binding to a surface oxygen with the monatomic adsorbates will form a dimer molecule whereas CO* and OH* can bind to form \ce{CO2}* and OOH* respectively. Incorporating these reactions in the dataset will allow for the exploration of intermediate surface reactions that are only possible on oxides.

Lastly, we also allowed for a four-fold rotational degree of freedom about the normal of the surface for all adsorbates. We randomly select the degree of rotation for each adsorbate on the surface after identifying the adsorbate sites.

\subsection{Structure Relaxation}

\begin{figure}[th!]
    \includegraphics[width=0.48\textwidth]{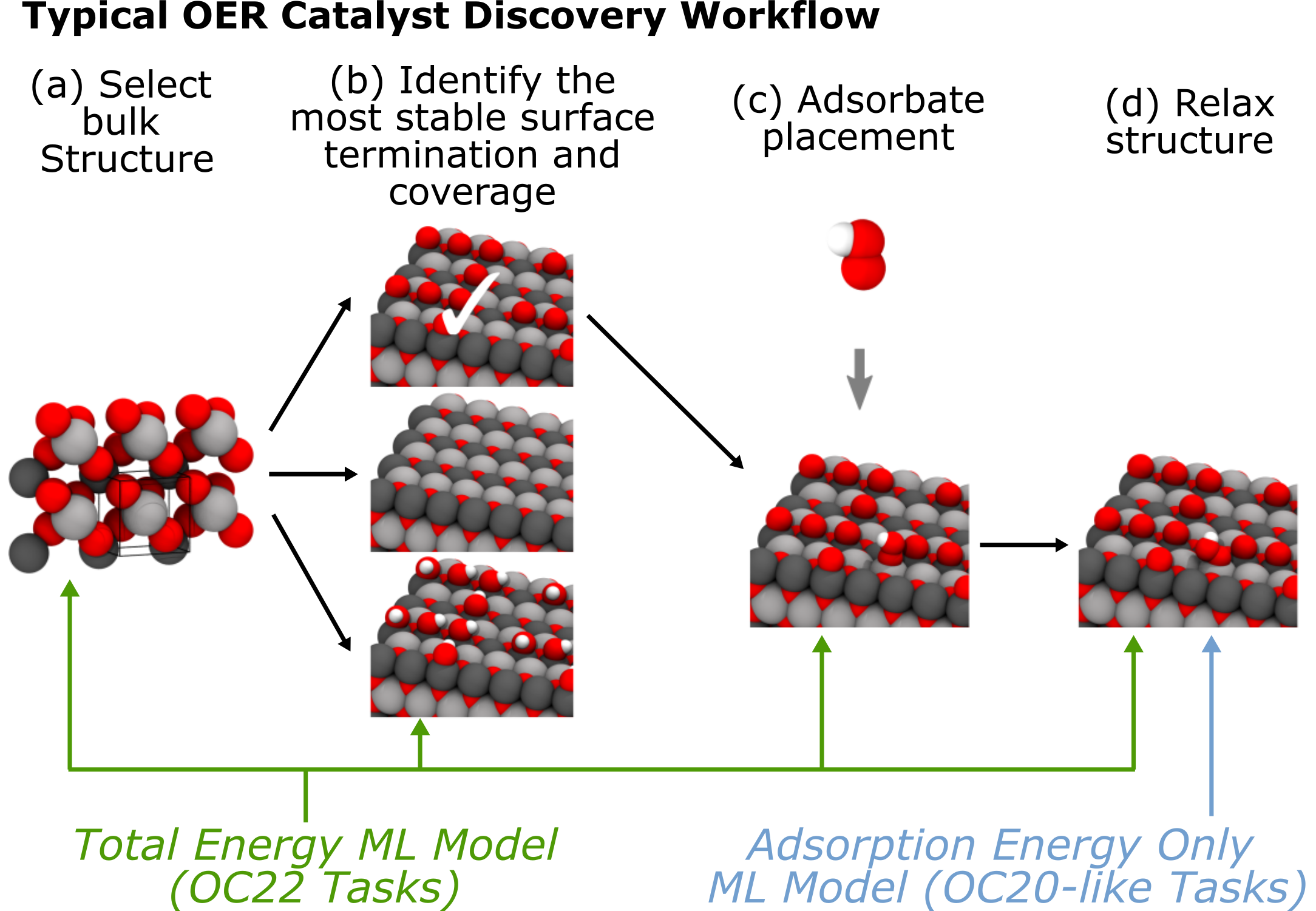}
    \caption{A typical \gls{OER} workflow, motivating the need for total energy models beyond adsorption energies. Total energy models would allow one to study all parts of this workflow, and not just the final relaxation like adsorption energy models. (a)  A bulk structure is selected from material datasets and a surface is created. (b) Surface terminations are enumerated and studied with \gls{DFT} to identify the most stable termination. Surface Pourbaix diagrams are created and used to make this decision. (c) Only after the most stable termination is identified, an adsorbate is placed and (d) The adsorbate+slab system is relaxed and the referenced adsorption energy is computed.}
    \label{fig:surfaces}
\end{figure}

The \gls{OC22} dataset uses different computational settings than those used for the \gls{OC20} dataset. The \gls{OC22} dataset models the exchange-correlation effects with the Perdew-Burke-Ernzerhof (PBE), generalized gradient approximation (GGA)~\cite{perdew1996generalized} which is generally accepted for modeling surface reactions on oxides\cite{Gonzalez2021, Heras-Domingo2019, VanDenBossche2017}. In contrast, the \gls{OC20} dataset utilizes the RPBE DFT functional. We also accounted for strong electron correlations in some transition metal oxides by applying the Hubbard U correction in accordance to the suggestions made by the Materials Project\cite{Jain2013}. The last row of Table~\ref{tab:comp_sample_table} shows the total number of slabs and adsorbate+slabs calculated using Hubbard U corrections. Although higher-level theory single-point calculations (e.g. hybrid functionals\cite{rousseau2020theoretical}) are often used to verify the final electronic structure and energy of a surface, they still use a scheme similar to the one here to obtain the optimized structure. Models developed for this dataset will greatly accelerate more accurate workflows by focusing expensive calculations on the most stable and relevant structures. 

In contrast to the \gls{OC20} dataset, all calculations were performed with spin-polarization to account for the significant spin states in metal oxides. Although some oxide materials exhibit magnetic polymorphism, we only considered one polymorph
for each slab with all slabs being initialized with ferromagnetic or nonmagnetic configurations in accordance to the magnetic moments of each metal suggested by~\citet{horton2019high}. These different magnetic states for a single crystal structure can significantly change thermodynamic properties at the surface. For example, rutile \ce{VO2} has been demonstrated to have several different spin states with nonmagnetic surfaces yielding significantly lower surface energies than ferromagnetic surfaces for the same slab\cite{Wahila2020}. For further details regarding the computational settings, we refer the reader to the \gls{SI}. 

We allowed all atoms of the slab and adsorbate+slab to be relaxed. This will not only yield a lower DFT energy, but also allows for more accurate calculations of the surface energy by ensuring both surfaces are relaxed. This is in contrast to the \gls{OC20} dataset where only the adsorbates and the surface atoms were relaxed.

Systems that did not converge ionically were set aside for use in alternative tasks. All intermediate structures, energies, and forces are stored for future training and evaluation. The algorithms implemented to produce all input slabs and adsorbate+slabs were constructed with the aid of Python Materials Genomics (pymatgen)\cite{Ong2013} and are available in the Open Catalyst Dataset repository (\url{https://github.com/Open-Catalyst-Project/Open-Catalyst-Dataset/tree/OC22_dataset}). All calculations are performed using the Vienna ab initio simulation package (VASP)~\cite{Kresse1994, Kresse1996, Kresse1996a, vasp-license, kresse1999ultrasoft}. In total, we used over 240 million core-hours to create this dataset.
\section{Tasks} 

The goal of the \gls{OC22} dataset is to efficiently simulate atomic systems with practical relevance to \gls{OER} and other oxide applications. One approach to screening materials relies on simple descriptors such as adsorption energy and surface energy. These descriptors alongside the Sabatier principle\cite{kuo2018measurements} and surface Pourbaix diagrams\cite{dickens2019electronic} can be used to correlate with more complex outputs like activity and selectivity. Unfortunately, the primary bottleneck to doing so are computationally expensive \gls{DFT} calculations. Calculations are further exacerbated for \gls{OC22} as its systems are larger and more complex than that of \gls{OC20}. Again, we focus on structure relaxations as they have been a useful means to informing catalyst activity for a broad range of applications\cite{huang2021adsorption, norskov2011density, bligaard2007ligand, hammer2000theoretical, seh2017combining, norskov2002universality}.
Models developed for \gls{OC20} have shown great progress on their proposed tasks\cite{gasteiger2021gemnet, ying2021transformers, shuaibi2021rotation, hu2021forcenet, gasteiger2022graph, oc20_perspective}. In all of the \gls{OC20} tasks, energies were referenced to represent adsorption energy. While advantageous for screening purposes, this referencing, however, implicitly limited models to only studying adsorbate+slab combinations and not any one in isolation. In the context of \gls{OER}, this is especially problematic as typical discovery pipelines require exploring different coverages and configurations of the surface \cite{gunasooriya2020analysis, back2020discovery, wang2020acid, patniboon2021acid, zagalskaya2021ab,vinogradova2018quantifying, flores2020active}. Figure \ref{fig:surfaces} illustrates a typical workflow for \gls{OER} where studying different surface terminations are necessary before running an adsorption calculation. Here, we propose modified variations of the \gls{OC20} tasks that would enable models to study surfaces with and without the presence of an adsorbate.

In all tasks, structures can contain a surface and adsorbate combination or just an isolated surface (a.k.a slab). The surface is defined by a unit cell periodic in all directions with a vacuum layer at least 12\text{\AA}. All ground truth targets are computed using \gls{DFT}.

We briefly summarize the \gls{OC20} tasks below. For all tasks, energy is referenced to correspond to adsorption energy. See the original \gls{OC20} manuscript for more details \cite{chanussot2021open}.
\textbf{\gls{S2EF}} takes a given structure and predicts the energy and per-atom forces.
\textbf{\gls{IS2RE}} takes an initial structure and predicts the relaxed energy.
\textbf{\gls{IS2RS}} takes an initial structure and predicts the relaxed structure. The size of the train and validation splits for each task is listed in Table~\ref{tab:splits}.

In the curation of both \gls{OC20} and \gls{OC22}, slabs and adsorbate+slabs were relaxed in parallel, with adsorbates being placed on unrelaxed slabs. \gls{OC20} makes an assumption in computing an adsorption energy such that the corresponding relaxed slab reference is comparable to that of the adsorbate+slab combination. This assumption was feasible given that the majority of the surface was constrained.

Unlike \gls{OC20} where surface atoms are constrained, all atoms in \gls{OC22} are unconstrained. While this enables the community to study other surface properties like surface energy, the assumption that the relaxed clean surface and adsorbate+slab surface are comparable no longer holds. Computing an adsorption energy in the same manner of \gls{OC20} would correspond to an incorrect reference, resulting in an ill-posed, noisy target (see \gls{SI} for more details). Instead, we modify the \gls{OC20} \gls{S2EF} and \gls{IS2RE} tasks to target DFT total energy rather than adsorption energy. We use the \gls{IS2RS} task as is with no modifications.

\textbf{\gls{RS2EF}} takes a given structure and predicts the DFT total energy and per-atoms forces. Compared to \gls{S2EF}, \gls{RS2EF} differs only in its energy prediction. \gls{S2EF} takes the DFT total energy and references it by subtracting off a clean surface and gas phase adsorbate energy. \gls{RS2EF} is only interested in the DFT total energy. The two tasks are related as follows:
\begin{equation}
\label{eq:rawads}
\hat{E}_{\gls{S2EF}} = \hat{E}_{\gls{RS2EF}} - E_{slab}^{DFT} - E_{gas}^{DFT}
\end{equation}

\textbf{\gls{RIS2RE}} takes a given structure and predicts the relaxed DFT total energy. Similar to \gls{RS2EF}, \gls{RIS2RE} is related to \gls{IS2RE} as follows:
\begin{equation}
\hat{E}_{\gls{IS2RE}} = \hat{E}_{\gls{RIS2RE}} - E_{slab}^{DFT} - E_{gas}^{DFT}
\end{equation}

\gls{DFT} total energies are not meaningful on their own. Physically relevant properties like adsorption energy include some reference. A model that can predict a \gls{DFT} total energy, however, gives the flexibility to reference to whatever is desired. Adsorption energy in this context would involve two predictions - one of the adsorbate+slab and one of the clean surface. For \gls{OER} this is particularly important to identify the most stable surface coverage (or termination). While this problem is also important for \gls{OC20}, those systems were much less complicated and the proposed adsorption energy tasks are typically sufficient. 

Of the proposed tasks, \gls{RS2EF} is the most general and closest to a \gls{DFT} surrogate. Models trained for this task would enable researchers to study properties derived from isolated surfaces such as surface stability with respect to the bulk energy (surface energy), a necessary and important step in the catalyst discovery pipeline. Total energies also allows us to leverage surface trajectories and their energies for training, data that was previously unusable in \gls{OC20} using the specified bare slab energy reference.
\begin{table*}
\caption{Size of train and validation splits. \gls{RS2EF} structures come from a superset of \gls{RIS2RE} systems, including unrelaxed systems (e.g. 50,810 train systems). Splits are sampled based on catalyst composition, ID for those from the same distribution as training, OOD for unseen catalyst compositions. Splits consist of both adsorbate+slab (adslabs) and slab systems. Validation and test splits are similar in size with exclusive compositions. }
\label{tab:splits}
\resizebox{\textwidth}{!}{%
\begin{tabular}{@{}ll|ccccccccc@{}}
\toprule
\multicolumn{2}{l}{\multirow{2}{*}{Task}} & \multicolumn{3}{c}{Train} & \multicolumn{3}{c}{ID} & \multicolumn{3}{c}{OOD} \\ \cmidrule(l){3-11} 
\multicolumn{2}{c}{} & \multicolumn{1}{c}{Adslabs} & Slabs & \multicolumn{1}{c|}{\textbf{Total}} & \multicolumn{1}{r}{Adslabs} & Slabs & \multicolumn{1}{l|}{\textbf{Total}} & \multicolumn{1}{r}{Adslabs} & Slabs & \textbf{Total} \\ \midrule
\multicolumn{2}{l|}{\gls{RS2EF}} & 6,642,168 & 1,583,125 & \multicolumn{1}{l|}{8,225,293} & 313,238 & 81,489 & \multicolumn{1}{l|}{394,727} & 356,633 & 94,036 & 450,669 \\
\multicolumn{2}{l|}{\gls{RIS2RE}} & 31,244 & 14,646 & \multicolumn{1}{l|}{45,890} & 1,701 & 923 & \multicolumn{1}{l|}{2,624} & 1,862 & 918 & 2,780 \\
\multicolumn{2}{l|}{\gls{IS2RS}} & 31,244 & 14,646 & \multicolumn{1}{l|}{45,890} & 1,701 & 923 & \multicolumn{1}{l|}{2,624} & 1,862 & 918 & 2,780 \\ \bottomrule
\end{tabular}%
}
\end{table*}
\section{Baseline GNN Models} 
A wide range of models for catalyst and molecular applications have been proposed \cite{gasteiger2021gemnet, gasteiger2022graph, shuaibi2021rotation, hu2021forcenet, ying2021transformers, godwin2021simple, liu2021spherical, sriram2022towards}. We evaluate our tasks using the latest state of the art models. Additionally, we baseline alternative model architectures including equivariant and (non)energy-conserving models. Code for all baseline models are implemented in PyTorch\cite{paszke2019pytorch} and PyTorch Geometric\cite{fey2019fast}, and are publicly available in our open source repository at \baselinesurl{}.

\gls{GNNs} have continued to grow in popularity as an efficient and accurate architecture for modeling atomic interactions. Unlike descriptor based models \cite{behler2016perspective, behlerparinello, chmiela2017machine, gap}, where hand crafted representations are used to describe atomic environments, \gls{GNNs} learn atomic representations through several message passing steps \cite{gilmermp}. Consistent with related work\cite{chanussot2021open,schutt2017schnet, klicpera2020directional}, graphs are constructed with atoms treated as nodes and interactions between atoms as edges. Periodic boundary conditions are accounted for in graph construction consistent with \gls{OC20}. A cutoff radius is introduced for computational tractability. 

We benchmark \gls{GNNs} that have either performed well on \gls{OC20} or other molecular datasets. For \gls{RS2EF}, we benchmark a larger sample of models including  SchNet\cite{schutt2017schnet}, DimeNet++\cite{klicpera2020fast},
ForceNet\cite{hu2021forcenet},
SpinConv\cite{shuaibi2021rotation}, PaiNN\cite{schutt2021equivariant}, GemNet-dT\cite{gasteiger2021gemnet}, and  GemNet-OC\cite{gasteiger2022graph}. \gls{IS2RS} baselines are limited to the top performing models - SpinConv, GemNet-dT, and GemNet-OC. \gls{RIS2RE} baselines include SchNet, PaiNN, DimeNet++, and GemNet-dT. Top performing \gls{RS2EF} models were also evaluated for \gls{RIS2RE} via an iterative relaxations approach\cite{chanussot2021open}. 

SchNet and DimeNet++ proposed continuous edge filters and directional message passing, respectively. ForceNet and SpinConv proposed architectures with direct force predictions in place of using energy derivatives with respect to atomic positions. PaiNN is an equivariant model with spherical harmonics up to order $l=1$. We modify PaiNN's original architecture to make direct force predictions as our experiments showed a boost in performance. GemNet-dT incorporates symmetric message passing, scaling factors, equivariant predictions, and several efficient architecture improvements over the similar DimeNet++. GemNet-OC expands on GemNet-dT to efficiently capture quadruplet interactions, the current state of the art model across all tasks for \gls{OC20}.

Unless otherwise noted, graph edges were computed on-the-fly via a nearest neighbor search for a cutoff radius of 6\text{\AA} and a maximum of 50 neighbors per atom. GemNet-OC uses different cutoffs for the type of interaction, e.g. triplets and quadruplets. Initial model sizes were taken directly from corresponding \gls{OC20} configurations. To accommodate for the fact \gls{OC22} has 16x less data, a light hyperparameter sweep was done for all models, with particular emphasis on learning rates, schedulers, and batch sizes. Effective batch sizes were set to $\sim$192-256 for \gls{S2EF} and $\sim$4-64 for \gls{IS2RE}. \gls{S2EF} models used identical learning rate schedulers to more fairly compare baselines, decaying the learning rate at epochs 2, 3, 4, 5, and 6. \gls{IS2RE} used a reduce on plateau learning rate scheduler. Full details on model hyperparameters and training configurations can be found in the \gls{SI}.

All experiments used the following loss function\cite{chanussot2021open} to balance energy and force predictions:

\begin{align}
    \mathcal{L}=\lambda_E\sum_i |E_i-E_i^{DFT}| \\ +\lambda_F\sum_{i,j}\frac{1}{3N_i} |F_{ij}-F_{ij}^{DFT}|^p \nonumber
\end{align}

where $\lambda_E$ and $\lambda_F$ are emperical parameters, $E_i$ is the energy of system $i$, $F_{ij}$ is the force on the $j$th atom in system $i$, $N_i$ is the number of atoms in system $i$, and $p$ is the norm order. With the exception of GemNet-dT and GemNet-OC which used $p=2$, all \gls{RS2EF} models used $p=1$. For \gls{RIS2RE} only the energy term is evaluated, i.e $\lambda_F = 0$. Baseline \gls{RS2EF} models were trained with $\lambda_E=1$ and $\lambda_F=N_{atoms}^2$ to insure size invaraince, as detailed by Batzner, et al. \cite{batzner20223, musaelian2022learning}
\section{Evaluation Metrics}
All our tasks use the same evaluation metrics proposed by \gls{OC20}. The only difference is rather than ground truth values being \gls{DFT} adsorption energies, we use \gls{DFT} total energies for \gls{OC22}. We briefly mention the metrics below but refer readers to the \gls{OC20} manuscript\cite{chanussot2021open} for a more detailed description.

\textbf{\gls{RS2EF}}: The \gls{RS2EF} task uses the same metrics as the \gls{OC20} \gls{S2EF} task. Metrics include Energy \gls{MAE}, Force \gls{MAE}, Force cosine, and \gls{EFwT}. Ground truth targets correspond to DFT total energy and per-atom forces.

\textbf{\gls{RIS2RE}}: Similarly, \gls{RIS2RE} uses the same metrics as the \gls{OC20} \gls{IS2RE} task. Metrics include Energy \gls{MAE} and \gls{EwT}. Ground truth targets correspond to the DFT total energy of the relaxed structure.

\textbf{\gls{IS2RS}}: \gls{IS2RS} metrics here are identical to that of \gls{OC20}. Metrics include \gls{ADwT}, \gls{FbT}, and \gls{AFbT}.  Ground truth targets are the relaxed structure. \gls{DFT} is also used to evaluate predicted relaxed structures.

Consistent with \gls{OC20}, our evaluation metrics still focus on accuracy. Given the complexity of \gls{OC22}, we are interested in how previously successful models will perform on larger more intricate systems. In addition, we focus on models that are significantly faster than traditional DFT-based techniques. Models that can calculate energy and force estimates in under 10ms would significantly aid oxide-related research.

\section{\gls{ML} Experiments}
The availability of large, diverse datasets like \gls{OC20} allows us to explore more interesting experiments alongside the \gls{OC22} dataset. In addition to training our baseline models on just \gls{OC22} we examine the extent the \gls{OC20} dataset and its pretrained models can benefit \gls{OC22} performance, and vice-versa.

\begin{figure}
    \centering
    \includegraphics[width=\columnwidth]{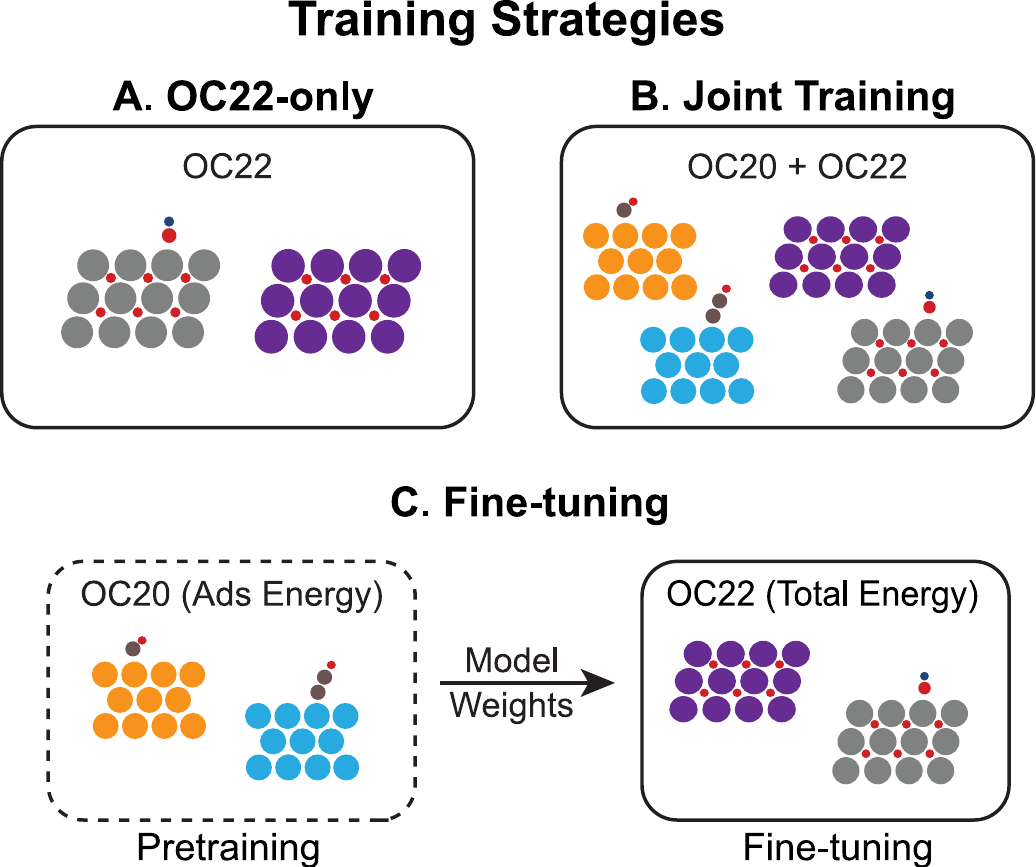}
    \caption{The various training strategies explored in \gls{OC22}. \textbf{A.} The \gls{OC22}-only strategy involves just using \gls{OC22} for the proposed tasks. \textbf{B.} Joint training refers to models trained on both \gls{OC20} and \gls{OC22} simultaneously. \textbf{C.} In fine-tuning, pretrained models for \gls{OC20} are used as starting points to train on just \gls{OC22}.}
    \label{fig:traing-expts}
\end{figure}

\begin{table*}[ht!]
\caption{Predicting total energy and force from a structure (\gls{RS2EF}). Results are shared for the \gls{OC22}-only, joint, and fine-tuning training strategies. Experiments are evaluated on the test set.}
\label{tab:rs2ef}
\resizebox{\textwidth}{!}{%
\begin{tabular}{@{}clllllllSSlSSlSSlS[round-precision=2]S[round-precision=2]l@{}}
& \multicolumn{19}{c}{\gls{RS2EF} Test}  \\
\cmidrule(r){1-19}
\mc{3}{l}{\mr{2}{*}{Training}} &
   &
  \mc{3}{l}{\mr{2}{*}{Model}} &
   &
  \mc{2}{c}{Energy MAE [eV] $\downarrow$} &
  \mc{1}{c}{} &
  \mc{2}{c}{Force MAE [eV/\AA] $\downarrow$} &
  \mc{1}{c}{} &
  \mc{2}{c}{Force Cosine $\uparrow$} &
  \mc{1}{c}{} &
  \mc{2}{c}{EFwT [\%] $\uparrow$} &
   \\ \cmidrule(lr){8-10} \cmidrule(lr){11-13} \cmidrule(lr){14-16} \cmidrule(lr){17-19}
\mc{3}{l}{} &
   &
  \mc{3}{c}{} &
   &
  {ID} &
  {OOD} &
  \mc{1}{c}{} &
  {ID} &
  {OOD} &
  \mc{1}{c}{} &
  {ID} &
  {OOD} &
  \mc{1}{c}{} &
  {ID} &
  {OOD} &
   \\ \cmidrule(r){1-19}
\mc{3}{l}{\mr{9}{*}{OC22-only}}      &  &
\mc{3}{l}{Median Baseline} &  & 163.4235 &	160.4547 &  & 0.0745 &	0.0729 &  & 0.0018 &	0.0018 &  & 0.00 &	0.00 &  \\
\mc{3}{c}{}                           &  &
\mc{3}{l}{SchNet~\cite{schutt2017schnet}} &  & 7.9241 &	7.9248 &  & 0.0601 &	0.0823 &  & 0.3633 &	0.2195 &  & 0.00 &	0.00 &  \\
\mc{3}{c}{}                           &  &
\mc{3}{l}{DimeNet$++$~\cite{klicpera2020directional,klicpera2020fast}} &  & 2.0952 &	2.4751 &  & 0.0426 &	0.0585 &  & 0.6062 &	0.4360 &  & 0.00 &	0.00 &  \\
\mc{3}{c}{}                           &  &
\mc{3}{l}{ForceNet~\cite{hu2021forcenet}} &  & {-} & {-} &  & 0.0564 &	0.0619 &  & 0.3507 &	0.2795 &  & 0.00 &	0.00 &  \\
\mc{3}{c}{}                           &  &
\mc{3}{l}{SpinConv~\cite{shuaibi2021rotation}} &  & 0.8364 &	1.9440 &  & 0.0377 &	0.0631 &  & 0.5905 &	0.4118 &  & 0.00 & 0.00 &  \\
\mc{3}{c}{}                           &  &
\mc{3}{l}{PaiNN~\cite{schutt2021equivariant}} &  & 0.9513 & 2.6300 &  & 0.0449	& 0.0583 &  & 0.4852	& 0.3449 &  & 0.00 &	0.00 &  \\
\mc{3}{c}{}                           &  &
\mc{3}{l}{GemNet-dT~\cite{gasteiger2021gemnet}} &  & 0.9385 &	1.2713 &  & 0.0316 &	0.0405 &  & 0.6647 &	0.5303 &  & 0.00 &	0.00 &  \\
\mc{3}{c}{}                           &  &
\mc{3}{l}{GemNet-OC~\cite{gasteiger2022graph}} &  & 0.3742 &	0.8287 &  & 0.0294 &	0.0397 &  & 0.6913 &	0.5500 &  & 0.02 &	0.00 &  \\
\midrule
\midrule
\mc{3}{l}{\mr{3}{*}{OC20-2M + OC22}}      &  &
\mc{3}{l}{PaiNN\cite{schutt2021equivariant}} &  & 0.3993 &	1.5293 &  & 0.0475 &	0.0644 &  & 0.4670 &	0.3201 &  & 0.01 &	0.00  \\
\mc{3}{c}{}                           &  &
\mc{3}{l}{SpinConv\cite{shuaibi2021rotation}} &  & 0.9310 &	1.7897 &  & 0.0361 &	0.0552 &  & 0.6210	& 0.4641 &  & 0.00 &	0.00 & \\
\mc{3}{c}{}                           &  &
\mc{3}{l}{GemNet-OC~\cite{gasteiger2022graph}} &  & 0.4207 &	0.9135 &  & 0.0288 &	0.0373 &  & 0.6929 &	0.5602 &  & 0.01 &	0.00 &  \\
\midrule
\mc{3}{l}{\mr{3}{*}{OC20-20M + OC22}}      &  &
\mc{3}{l}{PaiNN\cite{schutt2021equivariant}} &  & 0.3601 &	1.4538 &  & 0.0457 &	0.0605 &  & 0.4795 &	0.3410 &  & 0.01 &	0.00  \\
\mc{3}{c}{}                           &  &
\mc{3}{l}{SpinConv\cite{shuaibi2021rotation}} &  & 0.9721 &	1.5337 &  & 0.0355 &	0.0524 &  & 0.6008 &	0.4705 &  & 0.01 &	0.00 &  \\
\mc{3}{c}{}                           &  &
\mc{3}{l}{GemNet-OC~\cite{gasteiger2022graph}} &  & 0.3105 &	0.8274 &  & 0.0268 &	0.0366 &  & 0.7216	& 0.5845 &  & 0.08 &	0.01 &  \\
\midrule
\mc{3}{l}{\mr{2}{*}{OC20-All + OC22}}      &  &
\mc{3}{l}{SpinConv\cite{shuaibi2021rotation}} &  & 1.2968 &	1.7044 &  & 0.0397 &	0.0467 &  & 0.5293 &	0.4417 &  & 0.00 &	0.00 &  \\
\mc{3}{c}{}                           &  &
\mc{3}{l}{GemNet-OC~\cite{gasteiger2022graph}} &  & 0.3111 &	0.6893 &  & 0.0269 &	0.0342 &  & 0.7064 &	0.5859 &  & 0.07 &	0.00 &  \\
\midrule\midrule
\mc{3}{l}{\mr{4}{*}{OC20$\rightarrow$OC22}}      &  &
\mc{3}{l}{SpinConv\cite{shuaibi2021rotation}} &  & 1.1248 &	1.9661 &  & 0.0359 &	0.0505 &  &  0.6016 & 0.4578 &  & 0.00 & 0.00 \\
\mc{3}{c}{}                           &  &
\mc{3}{l}{GemNet-dT~\cite{gasteiger2021gemnet}} &  & 0.5715 &	1.0401 & &0.0307 &	0.0406 && 0.6734 &	0.5378 && 0.02 &	0.00 \\
\mc{3}{c}{}                           &  &
\mc{3}{l}{GemNet-OC~\cite{gasteiger2022graph}} &  & 0.2391 &	0.9378 &  & 0.0297 &	0.0407 &  & 0.6780 &	0.5364 &  & 0.13 &	0.00 &  \\
\mc{3}{c}{}                           &  &
\mc{3}{l}{GemNet-OC-Large*~\cite{gasteiger2022graph}} &  & 0.2173 &	1.0315 & &0.0272	& 0.0400 && 0.7298 &	0.5776 && 0.19 &	0.00 \\
\bottomrule
\multicolumn{19}{l}{*First fine-tuned energy output weights on \gls{OC22} energies, then fine-tuned entire network on \gls{OC22} energies+forces.}
\end{tabular}%
}
\end{table*}

The varied training strategies are summarized in Figure \ref{fig:traing-expts}. For each task we first study the performance using baseline models just trained on \gls{OC22} (\textbf{\gls{OC22}-only}). This is the standard strategy when introducing a new dataset.
Next, we leverage both \gls{OC20} and \gls{OC22} via \textbf{joint training} (\gls{OC20}+\gls{OC22}). In joint training we train a combined dataset of \gls{OC20} and \gls{OC22} systems. For \gls{RS2EF}, we explore combined datasets with different sizes of \gls{OC20} - 2M, 20M, and All. While the \gls{OC20} energies were originally expressed as adsorption energy, for these experiments we use the \gls{DFT} total energy which is also publicly accessible. 
One of the limitations to joint training is the need to train on a larger combined dataset, which can significantly increase training time. To address this, we additionally explore \textbf{fine-tuning} (\gls{OC20} $\rightarrow$ \gls{OC22}) experiments. In fine-tuning, models are initialized with pretrained weights learned from training on \gls{OC20}. The pretrained models are then fine-tuned by training on just \gls{OC22}. While approaches to fine-tuning vary in which portion of the network weights are updated, we limit our experiments to updating all the weights and leave more rigorous strategies as future work for the community \cite{kolluru2022transfer}. For \gls{RS2EF}, we experiment with fine-tuning using different fractions of the \gls{OC22} dataset. All fine-tuning experiments are performed using public \gls{OC20} adsorption-energy model checkpoints found at \modelurl.

Through these experiments we hope to share results that provide insights beyond just performance on \gls{OC22}. Building a dataset that spans all possible applications, chemical diversity, and level of DFT theory is not computationally feasible. However, as we demonstrate with \gls{OC22}, by leveraging large datasets, such as \gls{OC20}, we may be able to train effective models with much smaller datasets for specific domains; even if they contain critical differences like \gls{DFT} theory and material compositions.

\section{Results}
\begin{figure*}[th]
    \centering
    \includegraphics[width=0.9\textwidth]{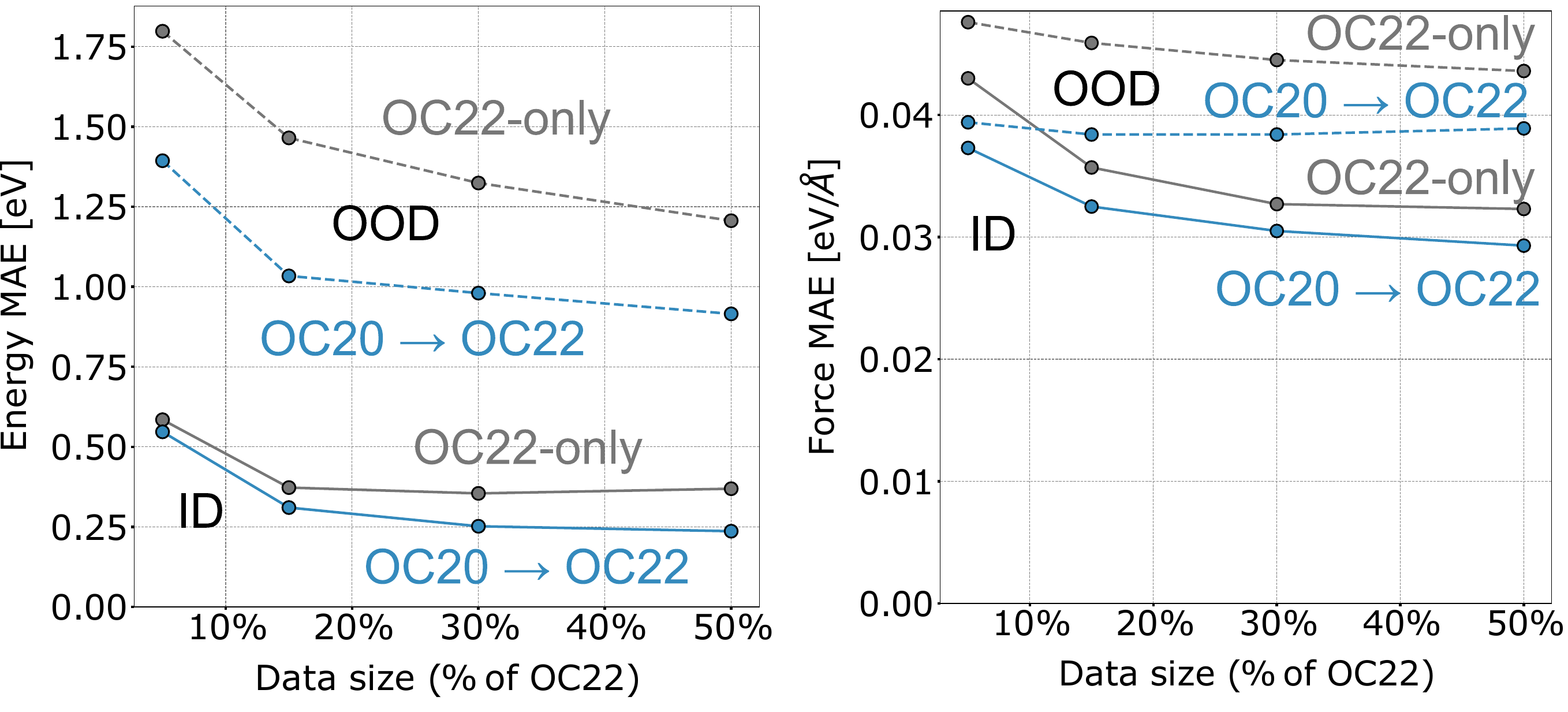}
    \caption{Results of GemNet-OC on \gls{RS2EF} across different training data sizes. Two strategies are compared here - \gls{OC22}-only and fine-tuning. Results are reported for both \gls{ID} (solid) and \gls{OOD} (dashed) on the test set.}
    \label{fig:goc-line}
\end{figure*}

\begin{figure*}[th!]
    \centering
    \includegraphics[width=\textwidth]{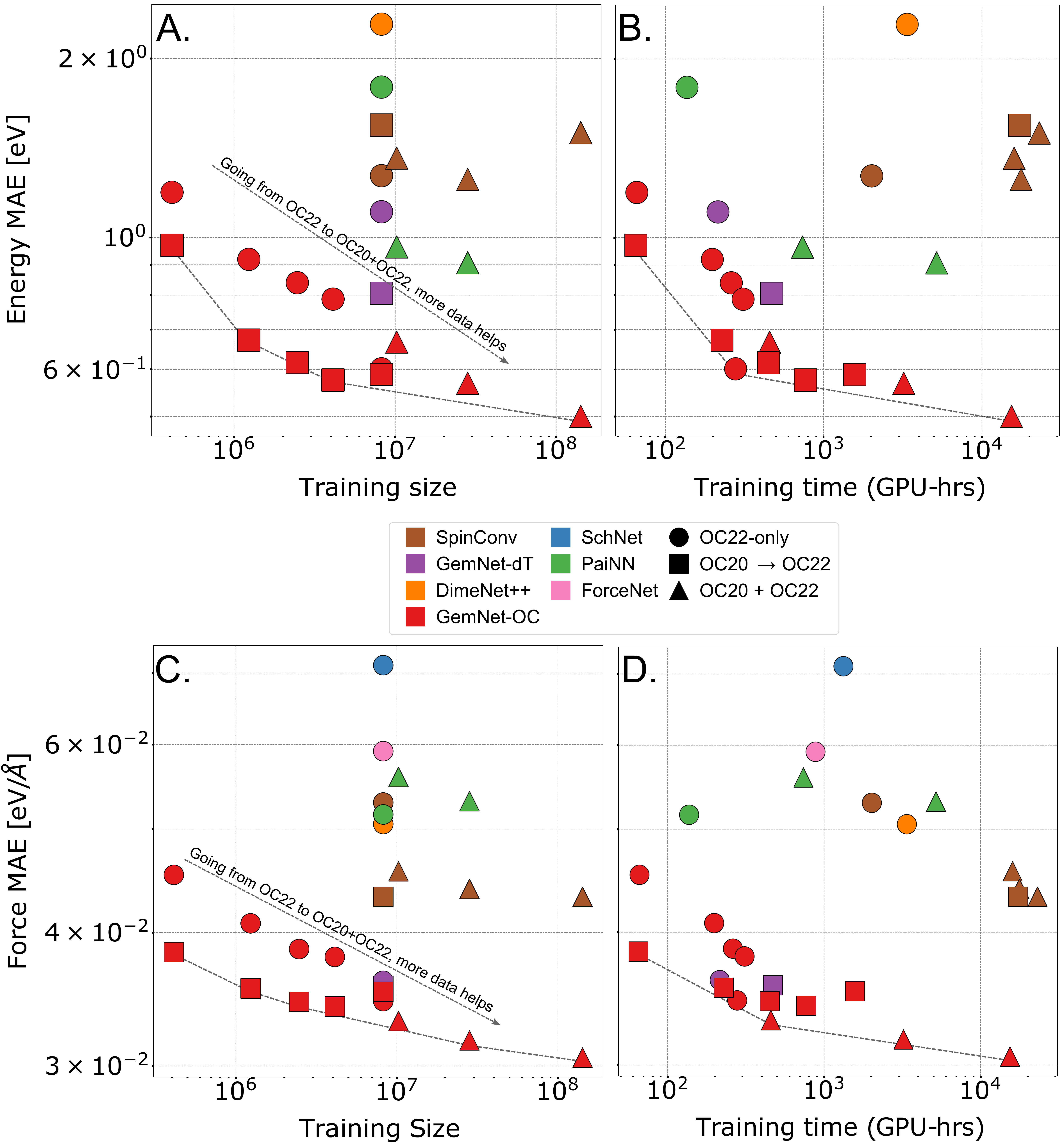}
    \caption{Summary of \gls{RS2EF} test results as a function of training size (A,C) and training time (B,D). Models are color coded and the respective training strategy is indicated by different shapes. For fixed dataset sizes, fine-tuning experiments see improvements in both energy and force predictions. Increasing data consistently helps performance when moving from \gls{OC22} to \gls{OC20}+\gls{OC22}. Pareto fronts are provided for current optimums across training sizes and times. Fine-tuning experiments do not consider the dataset sizes and training times used during pretraining. Results are averaged across both \gls{ID} and \gls{OOD} splits.}
    \label{fig:compute}
\end{figure*}

\begin{table*}
\caption{\gls{RS2EF} fine-tuning results trained on various fractions of the \gls{OC22} dataset. GemNet-OC\cite{gasteiger2022graph} was used for all experiments. Note, a fraction of $0\%$ for \gls{OC22} corresponds to the baseline of directly evaluating a pretrained checkpoint from \gls{OC20} on \gls{OC22}, with no additional training. All experiments are evaluated on the test set.}
\label{tab:rs2ef-ft}
\resizebox{0.8\textwidth}{!}{%
\begin{tabular}{@{}lcSSSSSSS[round-precision=2]S[round-precision=2]c@{}}
\multicolumn{10}{c}{\gls{RS2EF} Test}  \\
\toprule
\multirow{2}{*}{Training} & \multirow{2}{*}{Fraction of \gls{OC22}} & \multicolumn{2}{c}{Energy MAE [eV] $\downarrow$} & \multicolumn{2}{c}{Force MAE [eV/\AA] $\downarrow$} & \multicolumn{2}{c}{Force Cosine $\uparrow$} & \multicolumn{2}{c}{EFwT [\%] $\uparrow$} \\ \cmidrule(lr){3-4} \cmidrule(lr){5-6} \cmidrule(lr){7-8} \cmidrule(lr){9-10}   
 &   & {ID} & {OOD} & {ID} & {OOD} & {ID} & {OOD} & {ID} & {OOD} \\ \midrule
\multirow{5}{*}{OC22-only} 
& 5\% & 0.5847 &	1.7983 & 0.0430 &	0.0476 & 0.4967 &	0.4083 & 0.00 & 0.00 \\
& 15\% & 0.3727 &	1.4651 & 0.0357 &	0.0459 & 0.6142	& 0.4805 & 0.01 & 0.00 \\
& 30\% & 0.3546 &	1.3242 & 0.0327 &	0.0445 & 0.6589	& 0.5132 & 0.04 & 0.00 \\
& 50\%  & 0.3692 &	1.2064	  & 0.0323 &	0.0436 &  0.6566 &	0.5134 & 0.02 &	0.00 \\
& 100\% & 0.3742 &	0.8287	  & 0.0294 &	0.0396 &  0.6913 &	0.5500 & 0.02 &	0.00 \\ \midrule
\multirow{6}{*}{OC20$\rightarrow$OC22} 
& 0\% & 487.1210 & 434.6896 & 0.3650	& 0.3617 & 0.1944 &	0.1946 & 0.00 &	0.00 \\
& 5\%  & 0.5469 &	1.3938 & 0.0373	& 0.0394 & 0.5482 &	0.4766 & 0.00 & 0.00 \\
& 15\%  & 0.3103 &	1.0339 & 0.0325 &	0.0384 & 0.6212 &	0.5180 & 0.03 & 0.00 \\
& 30\% & 0.2519 &	0.9803 & 0.0305 &	0.0384 & 0.6565 &	0.5356 & 0.08 & 0.00 \\
& 50\%  & 0.2365 &	0.9154 & 0.0293	& 0.0389 & 0.6790 &	0.5464 & 0.13 & 0.01 \\ 
& 100\% & 0.2391 &	0.9378	  & 0.0297 &	0.0407 &  0.6780 &	0.5364 & 0.13 &	0.00 \\
 \bottomrule
\end{tabular}%
}
\end{table*}
We report results for all baseline models and tasks below. All validation results can be found in the \gls{SI}.

\begin{figure*}[t]
    \centering
    \includegraphics[width=\textwidth]{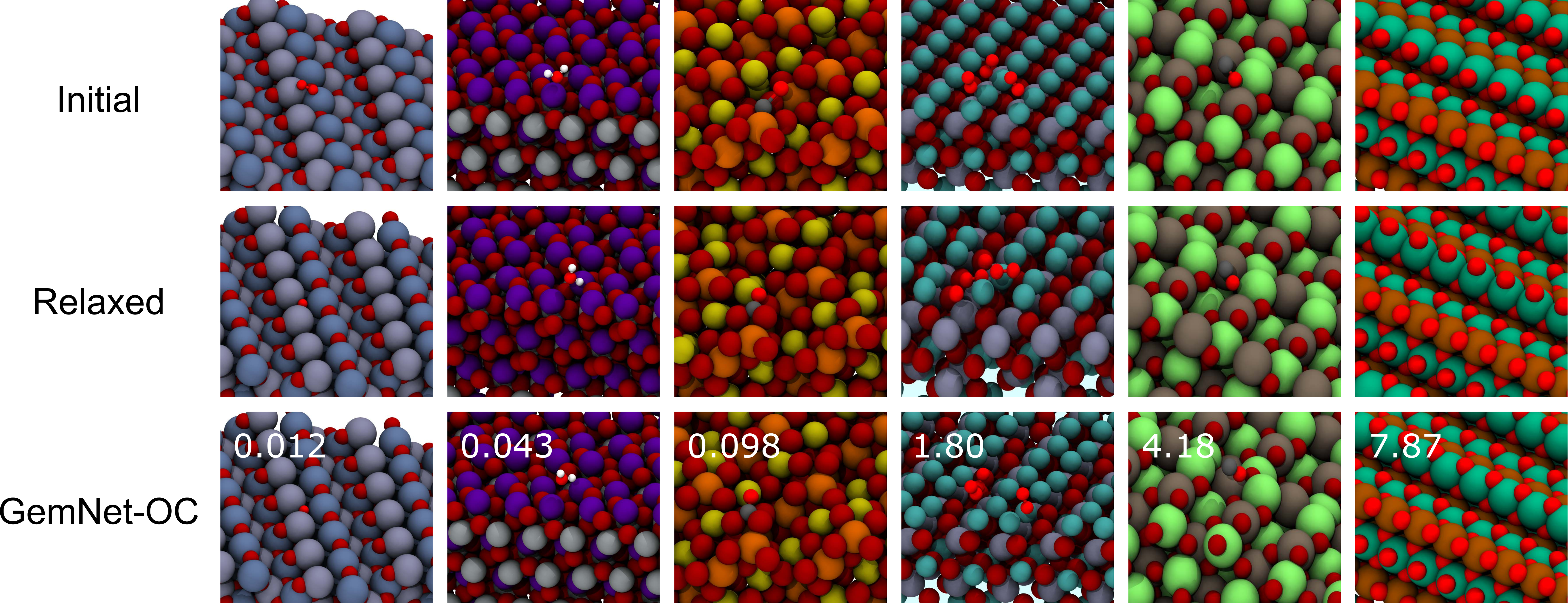}
    \caption{Demonstration of GemNet-OC solving the \gls{IS2RS} and \gls{RIS2RE} tasks via the relaxation approach. Initial, \gls{DFT} Relaxed, and the \gls{ML} predicted relaxed structures are shown for each system. The first three columns were randomly sampled from ``successful" cases in which \gls{RIS2RE} energy MAE was less than 0.1 eV, while the latter columns are ``failure" cases, with energy MAEs greater than 0.5 eV. Oxygen found in the adsorbate is illustrated with a high contrast red and made smaller to distinguish it from oxygen in the catalyst material.}
    \label{fig:mlrelax}
\end{figure*}

\textbf{\gls{RS2EF}}: Results on SchNet\cite{schutt2017schnet}, DimeNet++\cite{klicpera2020fast},
ForceNet\cite{hu2021forcenet},
SpinConv\cite{shuaibi2021rotation}, PaiNN\cite{schutt2021equivariant}, GemNet-dT\cite{gasteiger2021gemnet}, and  GemNet-OC\cite{gasteiger2022graph} are shown in Table \ref{tab:rs2ef} (top). All models make energy and per-atom force predictions. SchNet and DimeNet++ make force predictions via a gradient of energy with respect to atomic positions, while all other models make direct force predictions. Across all metrics, GemNet-OC performs the best. While GemNet-dT also demonstrates competitive force metrics, GemNet-OC significantly outperforms all models on energy based metrics. This may be due to the large receptive field (cutoff=12\AA) of GemNet-OC better capturing long-range interactions and its unique ability to explicitly capture quadruplet interactions. 

Results across the two test subsplits, In Domain (ID) and Out of Domain (OOD), are shown in Table \ref{tab:rs2ef}. As expected, \gls{ID} metrics are better than \gls{OOD}. Unlike \gls{OC20} where \gls{ID} and \gls{OOD}-based splits had fairly close metrics, \gls{OC22} \gls{OOD} metrics are substantially higher than \gls{ID}. By definition, \gls{OOD} contains combinations of material species not seen in the training set, i.e., if Ag-Cu is \gls{OOD}, then a Ag-Cu only interaction has never been seen during training. This suggests generalization in the context of total energy predictions is more challenging than a referenced adsorption energy.  
Although physically motivated, the \gls{OC20} adsorption energy target can also be thought of as a form of $\Delta$-learning \cite{shuaibi2020enabling, ramakrishnan2015big, zhu2019artificial}, simplifying the complexity of the problem to learning a correction to some base property. To explore this in the context of \gls{OC22}, we report results on a per-element linearly fit reference in the \gls{SI} that helps improve performance. We refrained from making this the base task for \gls{OC22} in order to encourage alternative schemes or approaches to target normalization. \gls{OC20} results on the proposed tasks are also available in the \gls{SI}, with similar poor performance suggesting \gls{RS2EF} to be a generally more challenging task.

\begin{table*}[ht!]
\caption{Predicting total relaxed energy from an initial structure (\gls{RIS2RE}). Results are shared for the \gls{OC22}-only, joint, and fine-tuning training strategies. Experiments are evaluated on the test set.}
\label{tab:ris2re}
\resizebox{0.7\textwidth}{!}{%
\begin{tabular}{@{}lllSSS[round-precision=2]S[round-precision=2]@{}}
\multicolumn{7}{c}{\gls{RIS2RE} Test}  \\
\toprule
\multirow{2}{*}{Approach} & \multirow{2}{*}{Training} & \multirow{2}{*}{Model} & \multicolumn{2}{c}{Energy MAE [eV] $\downarrow$} & \multicolumn{2}{c}{EwT [\%] $\uparrow$} \\ \cmidrule(lr){4-5} \cmidrule(lr){6-7}
 &  &  & {ID} & {OOD} & {ID} & {OOD} \\ \midrule
\multirow{10}{*}{Direct} & \multirow{4}{*}{OC22-only}
 & Median Baseline & 176.2560 &	171.8536 & 0.00	& 0.00 \\
 &  & SchNet & 2.0012 &	4.8468 & 1.03 &	0.45 \\
 &  & DimeNet++ & 1.9600 &	3.5186 & 0.65 &	0.38 \\
 &  & PaiNN & 1.7160 &	3.6835 & 0.88 &	0.38 \\
 &  & GemNet-dT & 1.6771 &	3.0837 & 1.49 &	0.45 \\ \cmidrule(l){2-7} 
 & \multirow{4}{*}{OC20+OC22} 
 & SchNet & 3.0384 &	4.2996 & 0.38 &	0.53 \\
 &  & DimeNet++ & 1.9614 &	3.4605 & 1.18 &	0.42 \\
 &  & PaiNN & 1.7330 &	3.7519 & 0.76 &	0.49 \\
 &  & GemNet-dT & 2.5234 &	4.2290 & 0.80 &	0.60 \\ \cmidrule(l){2-7} 
 & \multirow{1}{*}{OC20$\rightarrow$OC22} 
 & GemNet-OC*& 1.1527 &	1.7476 & 3.66 &	0.98 \\
 \midrule
 \midrule
\multirow{8}{*}{Relaxation} & \multirow{3}{*}{OC22-only}
 & SpinConv & 1.7367 &	2.6668 & 1.49 &	0.94 \\
 &  & GemNet-dT & 1.8129 &	2.0439 & 1.64 &0.83 \\
 &  & GemNet-OC & 1.3294 &	1.5841 & 2.02	& 1.40 \\ \cmidrule(l){2-7} 
 & \multirow{2}{*}{OC20+OC22} & SpinConv & 2.2959 &	2.5902 & 1.26 &	0.68 \\
 &  & GemNet-OC & 1.2007 &	1.5339 & 2.63 &	2.15 \\ \cmidrule(l){2-7} 
 & \multirow{3}{*}{OC20$\rightarrow$OC22} & SpinConv & 1.7997 &	2.8884 & 1.41 &	0.57 \\
 &  & GemNet-OC & 1.1201 &	1.8490 & 3.89 &	1.77 \\
  &  & GemNet-OC-Large & 1.2534 &	2.1154 & 1.60 &	0.98 \\
 \bottomrule
 \multicolumn{7}{l}{*GemNet-OC pretrained on \gls{OC20}+\gls{OC22} \gls{RS2EF}}  \\
\end{tabular}%
}
\end{table*}

Joint training experiments on \gls{OC20} and \gls{OC22} are conducted for the top performing models, GemNet-OC, PaiNN, and SpinConv. Table \ref{tab:rs2ef} additionally contains results of different sizes of \gls{OC20} combined with \gls{OC22}. To stay consistent with \gls{OC22}, DFT total energy targets were used for \gls{OC20}. With the addition of \gls{OC20} training data, GemNet-OC saw improvements in both energy and force predictions while PaiNN and SpinConv saw improvements to either only energy or forces, respectively. This suggests that despite the differences in DFT theory, the additional data is still meaningful in improving model predictions. However, increasing the amount of \gls{OC20} data had mixed results. GemNet-OC generally saw improvements across all metrics while SpinConv and PaiNN saw either minor improvements or worse performance. We note that training samples were randomly drawn, i.e., experiments with a larger proportion of \gls{OC20} would have seen fewer samples of \gls{OC22} during training. The differences in trends could be a result of model data efficiency and capacity. Exploring alternative sampling strategies to joint training could aid models and improve trends further. 
For our fine-tuning experiments, we evaluate GemNet-OC, SpinConv, and GemNet-dT models. Fine-tuning is performed
by first training a model on \gls{OC20}. This pre-trained model is then fine-tuned by training on
only \gls{OC22}.  Trained \gls{OC20} models are publicly available and were directly obtained from \baselinesurl{}. SpinConv saw little improvements on forces and worse performance for energies. GemNet-dT and GemNet-OC saw significant improvements to energy MAE and minor improvements to force MAE for \gls{ID} data. For \gls{OOD} data, GemNet-dT generally saw improvements with fine-tuning while all other models either saw similar or worse results. To drive performance further, we trained GemNet-OC-Large, a larger, more parameterized version of GemNet-OC under a more careful fine-tuning strategy. Here, the energy output weights were first fine-tuned on \gls{OC22} energies, afterwards, the entire network was fine-tuned on energies and forces. The large variant resulted in improved \gls{ID} energy and force predictions, with \gls{OOD} still seeing little or negative impacts. Fine-tuning experiments were extremely delicate and required careful selection of learning rates and hyperparameters, details are highlighted in the \gls{SI}. While our initial fine-tuning results were generally limited to energy improvements, we hope the future development of more rigorous methods could lead to better performance across all metrics. 

A potential benefit accompanying pretraining and fine-tuning is the need for less training data. A model initialized with meaningful weights could simplify the need to learn interactions and representations from scratch by utilizing an alternative dataset. To explore this, we evaluated the performance of a pretrained GemNet-OC model fine-tuned on various fractions of \gls{OC22}. As shown in Figure \ref{fig:goc-line}, a fine-tuned GemNet-OC consistently outperforms its \gls{OC22}-only variant across all data sizes for the \gls{ID} split, with diminishing returns for both strategies around $\sim$50\%. On \gls{OOD}, energy performance continues to improve with data size. In Table \ref{tab:rs2ef-ft}, we additionally show the performance of a pretrained \gls{OC20} GemNet-OC used to directly evaluate \gls{OC22} (Fraction = 0\%). As expected, energy metrics are extremely poor given the \gls{OC20} original target is adsorption energy. Force metrics are also extremely poor, suggesting the fine-tuning performance is not merely a result of a good pretrained model, but an actual transfer of knowledge from the two datasets. Figure \ref{fig:compute} illustrates the various models and approaches as a function of training size and time. Notably, we see a strong linear trend in performance with data size. With saturation yet in sight, we expect more joint dataset efforts to continue to aid in performance. While for a fixed dataset size, fine-tuning efforts improved performance, they were often more costly in training time (Figure \ref{fig:compute} B/D). We anticipate future fine-tuning developments to be not only more accurate, but efficient as well. Similar fine-tuning experiments with \gls{OC20} models trained on DFT total energy targets were also performed. Results were consistent with those shared above, suggesting that despite a difference in targets, models are learning a similar underlying representation that is being transferred to \gls{OC22}.

\textbf{\gls{RIS2RE}}: We explore two approaches for predicting relaxed energies from initial structures - ``Direct" and ``Relaxation" \cite{chanussot2021open}. The first directly predicts the relaxed energy with a single call to the model. The relaxation approach uses a \gls{RS2EF} model to run a structural relaxation - iteratively predicting forces and updating atomic positions until a relaxed structure and its corresponding energy is reached. While \gls{OC20} has shown relaxation based approaches to be superior to direct, they are 200-300x slower, motivating the potential benefits of direct models.

Table \ref{tab:ris2re} presents \gls{RIS2RE} results on both direct and relaxation approaches under the different training scenarios. Whereas \gls{OC20} saw relaxation based approaches to consistently perform better, we see mixed results here. The best relaxation-based approach, GemNet-OC, achieves an \gls{EwT} of 3.89\% indicating models have significant room for improvement. For the relaxation approach, fine-tuning consistently outperforms \gls{OC22}-only. The best direct approach, GemNet-OC, also only achieves an \gls{EwT} of 3.66\%. Here, joint training consistently hurts performance. Following literature efforts\cite{sriram2022towards}, fine-tuning was done from the top performing \gls{RS2EF} checkpoint - GemNet-OC \gls{OC20}-All+\gls{OC22}. While the best performing \gls{ID} results come from a direct approach, \gls{OOD} metrics are considerably better via the relaxation method, indicating their ability to better generalize. We evaluate \gls{OC20} \gls{RIS2RE} performance in the \gls{SI} and observe similar poor results, suggesting \gls{RIS2RE} to be a considerably more challenging variation.

\begin{table}
\caption{Predicting relaxed structures from initial structures (\gls{IS2RS}). All models predicted relaxed structures through an iterative relaxation approach. The initial structure was used as a naive baseline (IS baseline). Experiments are evaluated on the test set.}
\label{tab:is2rs}
\resizebox{0.5\textwidth}{!}{%
\begin{tabular}{@{}llS[round-precision=2]S[round-precision=2]S[round-precision=2]S[round-precision=2]S[round-precision=2]S[round-precision=2]S[round-precision=2]S[round-precision=2]@{}}
\multicolumn{8}{c}{\gls{IS2RS} Test}  \\
\toprule
\multirow{2}{*}{Training} & \multirow{2}{*}{Model} & \multicolumn{2}{c}{ADwT [\%] $\uparrow$} & \multicolumn{2}{c}{FbT [\%] $\uparrow$} & \multicolumn{2}{c}{AFbT [\%] $\uparrow$} \\ \cmidrule(lr){3-4} \cmidrule(lr){5-6} \cmidrule(lr){7-8}  
 &  & {ID} & {OOD} & {ID} & {OOD} & {ID} & {OOD} \\ \midrule
\multirow{4}{*}{OC22-only} 
 & IS baseline & 43.39 & 45.26  & 0.00 & 0.00  & 0.03 &	0.10  \\
 & SpinConv & 51.33 &	47.08  & 0.00  & 0.00  & 4.08 &	1.47  \\
 & GemNet-dT & 57.84 &	54.17  & 0.00 & 0.00 & 4.16 & 3.54  \\
 & GemNet-OC & 59.47 &	55.72 & 0.00 & 0.00  & 5.49 &	4.45  \\ \midrule
\multirow{2}{*}{OC20+OC22} 
 & SpinConv & 53.99 &	52.39  & 0.00  & 0.00 & 2.64 &2.38\\
 & GemNet-OC & 58.55 &	58.44 & 0.00 & 0.00 & 8.01 &	6.58  \\ \midrule
\multirow{3}{*}{OC20$\rightarrow$OC22} 
 & SpinConv & 54.21 & 51.42  & 0.08 &	0.00  & 6.31 & 3.24\\
 & GemNet-OC &  55.55 &	50.50  & 0.08 &	0.00  & 9.02 &	6.59 \\
  & GemNet-OC-Large & 57.23 & 54.63 & 0.00 & 0.00  & 10.41 & 8.09\\
 \bottomrule
\end{tabular}%
}
\end{table}
\textbf{\gls{IS2RS}}: To evaluate the prediction of relaxed structures from initial structures, we select the top performing \gls{RS2EF} models GemNet-dT, SpinConv, and GemNet-OC. Similar to \gls{OC20}, we use these models to run ML driven structure relaxations (Figure \ref{fig:mlrelax}). Relaxed structures were then evaluated with \gls{DFT} to determine whether the predicted relaxed structures are valid. Table \ref{tab:is2rs} shows GemNet-OC outperforming all other models across all metrics. Joint training and fine-tuning approaches both improve \gls{DFT} force based metrics over \gls{OC22}-only. GemNet-OC-Large fine-tuned achieves the best force metrics. Pursuant to \gls{OC20}, non-\gls{DFT} distance based metrics like \gls{ADwT} struggle to correlate well with the practical \gls{DFT} metrics \cite{oc20_perspective}. Both \gls{FbT} and \gls{AFbT} results indicate the models need significant improvement to achieve the level of accuracy needed for practical applications. 

\textbf{Does \gls{OC22} benefit \gls{OC20}?}
Alongside developing more accurate models, exploring augmentation strategies is another opportunity to improve performance on existing datasets like \gls{OC20}\cite{oc20_perspective}. An interesting question is whether \gls{OC22} data may improve model performance on \gls{OC20}. It has already been shown that the use of auxiliary data such as off-equilibrium MD or rattled data can lead to state-of-the-art results on \gls{OC20}\cite{gasteiger2022graph}. 

\begin{table}
\caption{GemNet-OC results trained on either \gls{OC20} or both \gls{OC20}+\gls{OC22} and evaluated on \gls{OC20} and \gls{OC22}. Results are averaged across all \gls{ID}/\gls{OOD} validation splits. Total energies are used for all dataset targets.}
\label{tab:oc20-res}
\resizebox{\columnwidth}{!}{%
\begin{tabular}{@{}lSSS@{}}
\toprule
Training & \multicolumn{1}{l}{Energy} & \multicolumn{1}{l}{Force} & \multicolumn{1}{l}{Force}\\
Data & \multicolumn{1}{l}{MAE [eV] $\downarrow$} & \multicolumn{1}{l}{MAE [eV/\AA] $\downarrow$} & \multicolumn{1}{l}{Cosine $\uparrow$} \\
\midrule
\multicolumn{4}{c}{OC22 evaluation} \\
OC20 & 55.9003 & 0.3842 & 0.1674 \\
OC20+OC22 & 0.6606 & 0.0305 & 0.6574 \\ \midrule
\multicolumn{4}{c}{OC20 evaluation} \\
OC20 & 0.3938 & 0.0219 & 0.6505 \\
OC20+OC22 & 0.3171 & 0.0231 & 0.6486\\
\bottomrule
\end{tabular}%
}
\end{table}

To that end, we explore the impact that jointly training with \gls{OC22} and \gls{OC20} has on \gls{OC20} performance.  Note \gls{OC22} is a significantly smaller and more limited dataset. \gls{OC20} contains $\sim$134M training data points and spans a large swath of materials. \gls{OC22} on the other hand is only $\sim$6\% of the size of \gls{OC20}, limited to only oxide materials, and places no constraints on atoms in the systems. Table \ref{tab:oc20-res} compares the performance of GemNet-OC trained on \gls{OC20} and \gls{OC20}+\gls{OC22} as evaluated on both \gls{OC20} and \gls{OC22} separately. As expected, when trained on only \gls{OC20}, \gls{OC22} metrics are poor - attributed to the the lack of oxides in \gls{OC20} and the difference in \gls{DFT} theories. When trained on \gls{OC20}+\gls{OC22}, however, we see a significant improvement in energy MAE ($\sim$20\%). Force based metrics are either no different or slightly worse. Despite the joint dataset containing only a small fraction of \gls{OC22}, it aided by a margin larger than any of the previous MD or rattled data efforts. Exploring in more detail as to how and why such improvements were observed could aid in systematically curating datasets to further improve \gls{OC20} performance.

\textbf{Adsorption Energy from Total Energy Models:}
\begin{table}
\caption{Adsorption energy predictions from total energy models on a subset of the OC22 validation ID dataset. Two scenarios are considered, \emph{Mixed-ML} where only the energy of adsorbate+slab is predicted with ML and \emph{Full-ML} where both the slab and adsorbate+slab are predicted with ML.}
\label{tab:ads_e_updated}
\resizebox{0.5\textwidth}{!}{%
\begin{tabular}{@{}llS[round-precision=3]S[round-precision=3]@{}}
\multicolumn{4}{c}{}  \\
\toprule
\multirow{1}{*}{Training} & \multirow{2}{*}{Model} & \multicolumn{1}{c}{Mixed-ML Ads. Energy} & \multicolumn{1}{c}{Full-ML Ads. Energy} \\
Data & & {MAE [eV] $\downarrow$} & {MAE [eV] $\downarrow$} \\ \midrule
OC22 & GemNet-OC & 0.678 & 0.767\\
OC20-All+OC22 & GemNet-OC & 0.691 & 0.724\\ \midrule
OC22 & PaiNN & 1.295 & 0.965\\
OC20-2M+OC22 & PaiNN & 0.795 & 0.825\\ \midrule
OC22 & SpinConv & 1.357 & 1.001\\
OC20-2M+OC22 & SpinConv & 0.984 & 0.980\\

 \bottomrule
\end{tabular}%
}
\end{table}

One of the difficulties with assessing the utility of total energy models is that, like \gls{DFT}, total energy values themselves are meaningless without an appropriate reference. Here we explore the performance of adsorption energy predictions via the total energy models. As previously detailed (see Tasks), calculating adsorption energies with \gls{OC22} data is ill-posed because of the potential inconsistencies resulting from relaxing the clean slab and adsorbate+slab in parallel. To address this, we sampled a subset of systems ($\sim$ 700) from \gls{OC22} and reran \gls{DFT} calculations with a more conventional procedure. Slabs were first relaxed, then adsorbates were placed on the relaxed slab, and finally the adsorbate+slab system was relaxed. Additionally, the bottom layers in the surface were fixed. This new data allowed us to validate the predicted adsorption energy from total energy models.

Total energy models were utilized to calculate adsorption energies using two different schemes. In the first approach, which we will refer to as \emph{Mixed-ML}, only the adsorbate+slab energy ($\hat{E}_{sys}$) in Equation~\ref{eq:mixed} is predicted with ML. Note, because \gls{OC22} includes systems with multiple adsorbates, the number of adsorbates, $n_{ad}$, is included in the calculation. Although this will yield the averaged adsorption energy across several adsorbates, for convenience, we will refer to it as the adsorption energy.
\begin{equation}
E^{Mixed}_{ad} = \frac{\hat{E}_{sys} - E^{DFT}_{slab} - E^{DFT}_{gas} \cdot n_{ad}}{n_{ad}}
\label{eq:mixed}
\end{equation}
Calculating the adsorbate+slab energy is the bottleneck in estimating adsorption energies, so replacing these DFT calculations with an ML potential could substantially improve the efficiency of screening new catalysts across a large swath of the chemical space. The second approach, \emph{Full-ML}, predicts the energy of the slab ($\hat{E}_{slab}$) and the adsorbate+slab ($\hat{E}_{sys}$), which would improve throughput even further.
\begin{equation}
E^{Full}_{ad} = \frac{\hat{E}_{sys} - \hat{E}_{slab} - E^{DFT}_{gas} \cdot n_{ad}}{n_{ad}}
\label{eq:full}
\end{equation}

The adsorption energy results for three models, GemNet-OC, PaiNN, and SpinConv are reported in  Table~\ref{tab:ads_e_updated}. All models were trained on the \emph{S2EF-total} task and used to run ML relaxations from the same initial structures as the DFT validation. More details on these calculations can be found in the \gls{SI}. As expected, the error is higher than predicting adsorption energy directly as was done in \gls{OC20}, but predicting total energy is a more challenging and general task.

Comparing the \emph{Mixed-ML} and \emph{Full-ML} adsorption energy results allows us to examine whether a cancellation or compounding of errors occurs. In the \emph{Full-ML} approach, the ML potential could have some systemic bias compared to the total energy DFT labels, yet still produce accurate adsorption energies if this error is ``cancelled out" by subtracting off the ML slab prediction containing a similar bias. Alternatively, because ML is used for both an adsorbate+slab and slab prediction, errors could accumulate in the final prediction. While GemNet-OC does not observe any cancellation of errors, PaiNN and SpinConv see some improvements on the \emph{Full-ML} approach. However, these trends go away when the training is augmented with \gls{OC20} data, with both approaches producing similar results. While adsorption energy prediction from total energy models is still far from chemical accuracy $\sim$ 0.1 eV, we demonstrate the utility of current models compared to literature studies below. Additionally, we envision that the release of the \gls{OC22} dataset will lead to rapid modeling improvements as was the case for \gls{OC20}. 

\textbf{Comparisons with Literature:} To demonstrate relevance of our models beyond the \gls{OC22} dataset, we use our models to compare predicted adsorption energies and trends with corresponding existing data in the literature for \ce{O^*}, \ce{H^*}, and \ce{OH^*}.

\begin{figure*}[th!]
    \centering
    \includegraphics[width=0.9\textwidth]{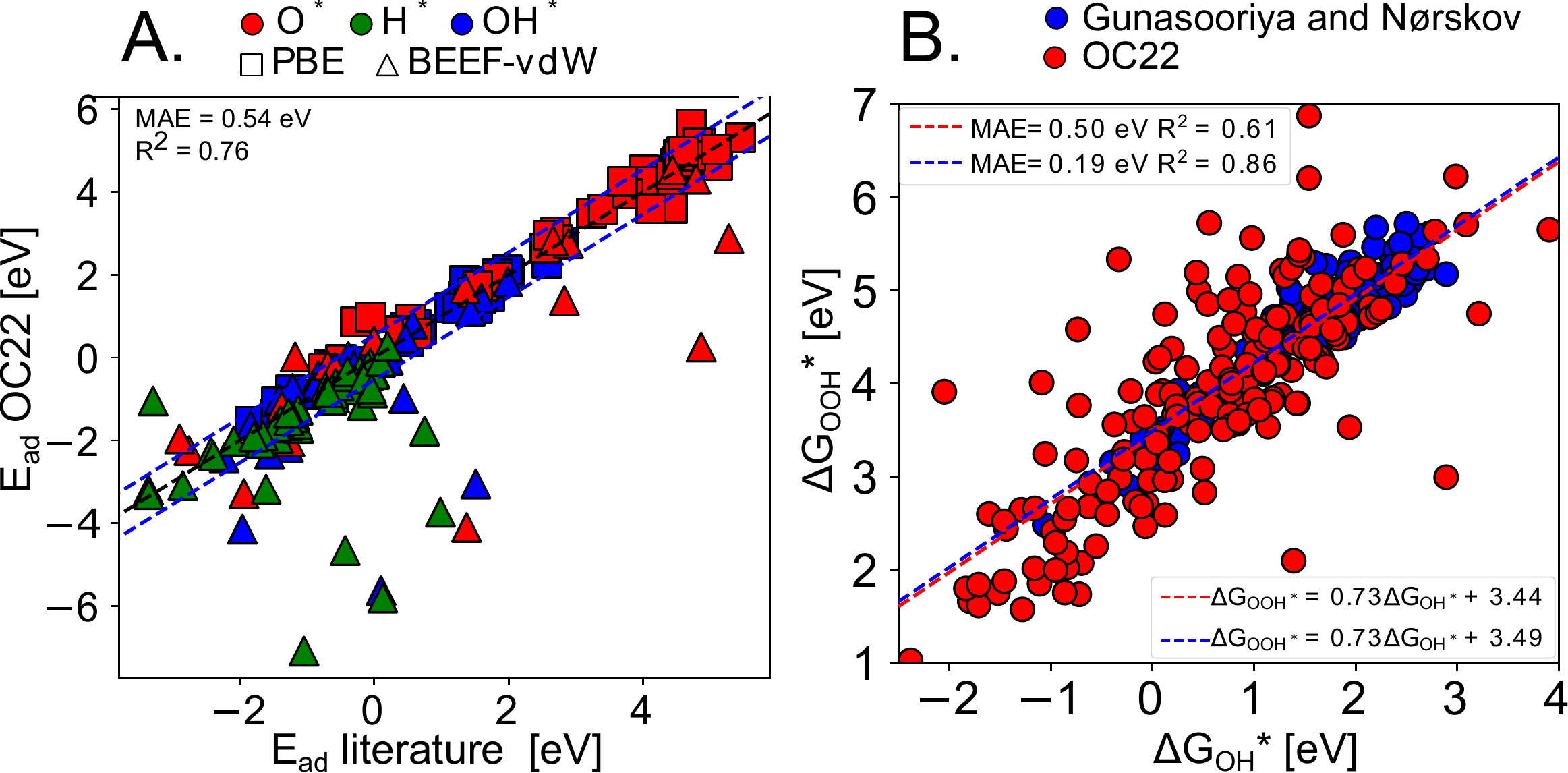}
    \caption{(A) Comparison of OC22 predicted (y-axis) and literature (x-axis) values for the adsorption energies of \ce{O^*}, \ce{H^*}, and \ce{OH^*} across different OOD metal oxide compounds for rutile structures (see SI for a comparison of perovskite structures). A parity line (black-dashed) is provided for reference as well as a line above and below to indicate the mean absolute error (blue-dashed). (B) A comparison of $\Delta G_{OOH*}$ (y-axis) and $\Delta G_{OH*}$ (x-axis) with predicted (red) and literature (blue) data points shown along with their corresponding linear fits (dashed lines). All predictions were performed using the GemNet-OC OC20+OC22 model.}
    \label{fig:lit_compare}
\end{figure*}

Figure~\ref{fig:lit_compare}(A) plots our predicted values for the adsorption energies of \ce{O^*}, \ce{H^*}, and \ce{OH^*} against a sample set of literature values\cite{dickens2019electronic, comer2018analysis} for rutile surfaces. The adsorption energies demonstrated a linear correlation of at least 75\% with the majority of predicted values within 0.6 eV of the literature. The 1 eV discrepancy of several outliers could be accounted for by several differences in computational parameters between the literature values that utilized BEEF-vdW\cite{dickens2019electronic} and the training data for OC22. These discrepancies include the lack of Hubbard U corrections and the absence of spin-polarization in the literature. Despite the large deviation in some of the BEEF-vdW data points, this comparison demonstrates that a majority of our predicted adsorption energies for some compositions are agnostic of the functional. 

From the adsorption energy, we can obtain scaling relationships which are useful for identifying optimal catalysts across a variety of materials. Predicting these trends with OC22 will demonstrate the value of our model as a viable tool for screening catalysts. Figure~\ref{fig:lit_compare}(B) shows the scaling relationship between the Gibbs adsorption energy of \ce{OH^*} and \ce{OOH^*} calculated at standard conditions (T=298.15~K, P=1~bar) in literature\cite{gunasooriya2020analysis} and predicted data points using OC22 (see Table S11 in the SI for the Gibbs energy corrections). Our linear fitting of the slope (0.73) and intercepts (3.44) for $\Delta G_{OOH*}$ vs $\Delta G_{OH*}$ are consistent within 0.05 eV of the literature slope (0.73) and intercepts (3.49)\cite{gunasooriya2020analysis}. We also demonstrated a similar fitting for $\Delta G_{O*}$ vs $\Delta G_{OH*}$ (see SI) albeit with our intercept overestimating by approximately 0.7 eV. Despite the significantly higher MAE and lower $R^2$ in the predicted relationships, we were still able to obtain a significant linear correlation above 60\%.

While adsorption energy plays an important role for catalyst discovery, incorporating vibrational and zero point energy contributions are necessary for Gibbs energy calculations. Gibbs energy of adsorption ($\Delta G_{ad}$) is often necessary for constructing accurate reaction pathways, creating microkinetic models, and determining the overpotential. However, at specific atmospheric conditions, the adsorption energies can generally be shifted by a constant correction value to obtain $\Delta G_{ad}$. To demonstrate this, we plotted a minimal set of datapoints (15 - 20) of $\Delta G_{ad}$ at standard temperature and pressure against the adsorption energy to obtain a constant shift ($\Delta G_{corr}$) between the two quantities. We can therefore add $\Delta G_{corr}$ to the adsorption energy of any system with the same adsorbate to get $\Delta G_{ad}$ (see Figure S4). This method of calculating $\Delta G_{ad}$ when applied to predicted adsorption energies circumvents the need for a separate model of $\Delta G_{ad}$.

For more details in regards to literature validation and other scaling relationships, we refer the reader to the SI.
\section{Discussion} 

There are many challenges to building large datasets and fitting generalizable models in computational catalysis, some of which were recently summarized \cite{oc20_perspective}. All of the challenges described also apply to the \gls{OC22} dataset - model performance varies across adsorbates and materials, direct force predictions tend to perform the best despite breaking energy conservation, developing helpful metrics for common tasks like local relaxations is difficult, and choosing the right calculations to improve the performance and generalizability of models is challenging. This work adds to these difficulties by highlighting additional challenges in capturing long-range interactions, developing models that go beyond adsorption energy, and fitting models with multiple datasets and levels of theory. 

The performance of baseline models in this work is impressive given the difficulty of predicting the total system energy of complex oxide surfaces, but challenges still remain. The best results on the most general \gls{RS2EF} task using a transfer learning approach from OC20 has an energy MAE of 0.24 eV for ID performance and 0.94 eV for OOD performance. Using that same model to predict relaxed total energies yields energy MAEs of 1.12 eV for ID and 1.85 eV for OOD predictions. These results are somewhat more impressive on a per-atom basis as is common for formation energy estimates of materials. However, for predicting experimentally-relevant properties like the overpotential for the \gls{OER}, these results are far from sufficient. We note that the initial baseline models for \gls{OC20} were similarly unhelpful for catalyst activity predictions, but rapid contributions from the broader community greatly improved their accuracy and predictive power. We hope that similar progress is seen for the tasks here. We also expect that the current models may already be helpful for certain more limited tasks, such as accelerating future oxide calculations with the use of online fine-tuning \cite{musielewicz2022finetuna}.

The tasks proposed in this work aim to push the community more in the direction of a general purpose potential, rather than separate models for each specific property. As an example, the tasks in \gls{OC20} were limited to the prediction of a specific property - the adsorption energy, following the most common approach in the community\cite{Sorescu2002, Sholl2009, Wellendorff2015}. This was a reasonable choice as the adsorption energy is a common descriptor for catalytic properties, and the adsorption energy itself was thought to be easier to fit than the DFT total energies. However, defining the tasks in this way meant that resulting models could only predict the adsorption energy and were unhelpful for predicting other surface properties like the surface energy. These limitations are highlighted in oxide catalysis where the stability of various surface terminations is needed. The total energy tasks in this work should encourage models that serve as general \gls{DFT} surrogates - making predictions on a much wider range of properties. 

\subsection{Future directions}

\textbf{Long-range interactions}: The \gls{OC22} dataset contains long-range interactions that are likely difficult to capture in existing \gls{GNN} models. Unlike metal surfaces which have a sea of electrons that can screen interactions, many of the oxides in \gls{OC22} are semiconductors with considerable partial charges (especially on the oxygen atoms). Electrostatics have very long range effects (energy decaying as $1/r$), and the partial charges can vary from system to system. The interaction of magnetic spins in systems with spin polarization is also long-ranged. This poses a challenge for the \gls{GNN}s used in this work, which are often developed under the assumption that local interactions dominate. The use of several message passing steps or long-range local cutoffs may allow for these long-range interactions to be captured. There has been considerable effort in developing \gls{ML} models that include long range interactions\cite{ko2021general, zubatiuk2021machine, behler2021machine}, and we expect those approaches to be very useful in improving predictions for \gls{OC22}.

\textbf{Higher-level theory}: The \gls{OC22} dataset also highlights the challenges of requiring multiple levels of theory for varying properties and materials. The \gls{OC20} dataset was constructed with the RPBE functional and neglected spin polarization, which represented a good trade-off between accuracy and computational cost for adsorption energies. However, some oxide surfaces require proper selection of Hubbard U corrections and can exhibit significant spin polarization. Combining datasets with multiple levels of theory, or upgrading datasets from less accurate to more accurate methods are popular questions in the small molecule community \cite{grambow2019accurate, duan2022two}, but applying these ideas to \gls{OC20}/\gls{OC22} will require extending these approaches to large datasets and inorganic materials, and we hope the community rises to this challenge. An obvious future direction is to improve the data quality with far more expensive hybrid functionals on the relaxed structures here. 

\textbf{Magnetic and charge effects}: While additional information beyond just atom positions and atomic numbers have yet to improve performance on earlier datasets like \gls{OC20}, it remains an open problem, how to best incorporate additional physics. In the context of \gls{OC22}, magnetic configurations play an important role in oxide chemistry. Oxides exhibit different magnetic configurations for the same structure. These magnetic polymorphs can lead to different energetic, structural and magnetic trajectories along with different oxidation states during relaxation which can drastically affect chemisorption\cite{biz2020magnetism, biz2021strongly, ren2021spin}. Identifying ways to include magnetic moments in future models or training strategies might be an important contribution to improving GNN performance.

We would also like to explore the effect of charge balancing in the future for our oxide systems. In contrast to metallic systems, semiconductors can thermodynamically favor surface reconstruction over electron promotion when dopants, charges or stoichiometric defects are present. Essentially, bonds can be broken or created in such a way that electrons are prevented from getting promoted into the conduction band in a process called self-compensation \cite{zunger2003practical, pashley1989electron}. As a result, nominally identical initial structures, but with different numbers of electrons (and consequently, Fermi levels), can relax to significantly different final geometries (e.g., adatoms being expelled or dimers formed or broken \cite{voznyy2013dynamic}). This effect can be extremely long-range as it only depends on the total number of shared electrons in the system, i.e., a vacancy introduced on one side of the slab may affect a local geometry on the other side.

Capturing the complexity of such phenomena with ML models would require a more in-depth analysis of a system's Fermi level's and the model's ability to capture long-range interactions. In particular, the differences in forces and in final geometries induced by electronic doping should be explored, as well as the ability of GNNs to differentiate between them using an additional doping descriptor. To disentangle the effect of oxygen vacancies on Fermi level vs. their local effects, a purely charging-based dataset may be prepared. While Fermi level positions are results of DFT calculations, electron counting and band filling\cite{pashley1989electron, voznyy2012charge}, can be leveraged to provide empirically similar information as additional data.


\textbf{Solvation effects}: The impact of solvation is another area for future research. Although we did not directly model solvation effects in this work, we are able to account for the consequences of partial surface dissolution by incorporating random oxygen vacancies at varying coverages. Oxide surfaces are prone to partial dissolution from solvents which can lead to surface vacancy defects that can modify catalytic properties\cite{HerasDomingo2019}. Modeling OER on \ce{RuO2} and \ce{IrO2} in the presence of these defects allowed for previous computational studies to obtain descriptors of overpotential and activity that more closely reflect experimental observations\cite{Dickens2017, Zagalskaya2020}.

\textbf{Training strategies}: Joint training on both the \gls{OC20} and \gls{OC22} datasets leads to several unexpected results. Surprisingly, naively fitting on both \gls{OC20} and \gls{OC22} (much smaller dataset) leads to large accuracy improvements for predicting \gls{OC20} energies, as shown in Table \ref{tab:oc20-res}. In addition,  models trained on either \gls{OC22} or \gls{OC20}+\gls{OC22} both appear to follow the same log-log scaling for energy MAE (Figure \ref{fig:compute}). These observations  open the door to using a wide array of existing large datasets (NOMAD\cite{nomad}, Materials Project\cite{Jain2013}, OQMD\cite{saal2013materials}) that although different, could aid in model development. These ideas can be rationalized if all of these datasets together can help learn more flexible and useful representations, regardless of their specific tasks or details. 

Fine-tuning and transfer learning baselines were investigated as potential routes to improve accuracy across both \gls{OC20} and \gls{OC22} and reduce the computational intensity of training \gls{GNN}s for these tasks. The most accurate models for both OC20 and OC22 were models trained on both datasets simultaneously, which indicates that a common representation can be learned and shared by both datasets. Surprisingly, the limited fine-tuning experiments in this work did not improve substantially on the accuracy/cost Pareto front (Table \ref{fig:compute}). However, there are many possible fine-tuning strategies and a large number of variations (e.g. which sections of the \gls{GNN} to freeze or fit, or leaving this decision to an attention block \cite{kolluru2022transfer}), and we expect more progress from the community in this area. These approaches are necessary to encourage the re-use of large models, and to reduce the computational cost of obtaining state-of-the-art models for future small datasets. 

\textbf{Alternative property predictions}: Models trained on \gls{OC22} could predict the total energies for any slab or adsorbate+slab which ultimately allows us to determine any thermodynamic quantity including adsorption energy, surface energy, and reaction energy. Adsorption and reaction energies are useful for identifying viable catalysts. We can also predict the surface energy in order to construct elaborate phase diagrams which can be used to assess the thermodynamic stability of a surface at varying adsorbate coverages. Pourbaix diagrams (applied potential vs pH) are especially important for determining the thermodynamic viability of electrocatalysts. The surface energy can also be used to model the equilibrium crystal structure or Wulff shape. With a predictive model that circumvents DFT calculations, all these applications, which ordinarily require hundreds of DFT calculations, are possible with little to no computational cost.

This dataset will have a broad impact in discovering oxide catalysts for a variety of reaction families and unraveling complex reaction mechanisms in these systems. Oxide materials are likely present in any reaction under strong oxidative conditions, such as the accelerated degradation of long-lived contaminants like PFOA\cite{liang2018electrochemical} or systematically upgrading chemical building blocks \cite{vedrine2019metal}. Photocatalysis, which directly uses available sunlight to drive chemical reactions also relies heavily on oxides such as TiO$_2$ due to their desirable optical properties \cite{comer2018analysis} and could benefit from this dataset. One example which is currently computationally expensive to study is the \gls{MvK} mechanism, which is one of the most common catalytic mechanisms in ionic crystals\cite{Mars1954, Hinuma2018}. In the \gls{MvK}, an adsorbate binds to a surface oxygen to form a new intermediate which desorbs to leave behind an oxygen vacancy, which can later be replenished by oxygen atoms from incoming adsorbates. By explicitly including oxygen defects and vacancies in the dataset generation process, we hope the resulting models will be helpful for accelerating these studies. Similar reactions that could benefit from these approaches are \ce{CO2} capture on carbides \cite{gracia2009mars} or nitrate reduction on nitrides\cite{abghoui2017electrochemical}.

\textbf{Experimental outlook}: Ultimately, the goal of developing accurate computational models is to inform experimental design and discovery, either through direct quantitative agreement or by providing insight into the underlying physical phenomena. It is important to be cognizant of the fact that the catalyst systems we simulate with DFT are idealized versions of the actual physical systems observed in experiments. For instance, the structure and composition of an oxide catalyst are prone to change during the reaction due to interactions with reaction intermediates and the surrounding medium, which complicates the connection with idealized DFT calculations. One way to enhance this work would be to complement it with experimental validation and auxiliary data from other modeling techniques and experiments like microkinetic models and operando spectroscopy. In spite of these challenges, DFT has proven to be an essential tool for providing atomistic level insights for catalysis and we envision that the improvements made in modeling oxide catalysts as a result of the \gls{OC22} dataset, e.g. considering the influence of surface coverages and defects on the oxide structure at a much larger scale than previously possible, will pave the way towards strengthening the connection with experiments and unearthing underlying catalyst design principles.

\begin{suppinfo}

The supporting information contains details on \gls{OC20} \gls{RS2EF} and \gls{RIS2RE} results, results using an alternative reference scheme, a discussion on adsorption energy for \gls{OC22}, performance on \gls{OC22} adsorbate+slabs and slabs, independently, training and hyperparameters for baseline models, full results on the validation splits, additional validation with literature, a description of calculated corrections for Gibbs adsorption energy, and the Hubbard U corrections used. The full dataset is provided at \ocpurl{} and available in an ASE\cite{ase_Hjorth_Larsen_2017} trajectory or model-ready LMDB format. Baseline models, dataloaders, and trainers are provided in the open source repository \baselinesurl{}.

\end{suppinfo}

\begin{acknowledgement}

The authors acknowledge Shaama M. Sharada, Samira Siahrostami, Andrew J. Medford, Tiago F. Goncalves, Selin B. Bilgi, and Sophia Kassabian for their assistance in reviewing calculations and helpful discussions. The authors also acknowledge helpful discussions with Aleksandra Vojvodic, John Kitchin, as well as Johannes Gasteiger on modeling choices.

\end{acknowledgement}
\bibliography{main_arxiv}

\begin{mcitethebibliography}{116}
\bibitem[vas()]{vasp-license}
``The calculations in this work have been performed using the ab-initio
  total-energy and molecular- dynamics package VASP (Vienna ab-initio
  simulation package) developed at the Institut f{\"u}r Materialphysik of the
  Universi{\"a}t Wien". \relax
\mciteBstWouldAddEndPunctfalse
\mciteSetBstMidEndSepPunct{\mcitedefaultmidpunct}
{}{\mcitedefaultseppunct}\relax
\EndOfBibitem
\end{mcitethebibliography}

\clearpage
\onecolumn
\section{\gls{OC20} \gls{RS2EF} and \gls{RIS2RE} results}
To enable the comparison of total energy metrics between the \gls{OC22} and \gls{OC20} datasets, we trained baseline \gls{OC20} models for the proposed \gls{RS2EF} and \gls{RIS2RE} tasks. Table \ref{tab:oc20-rs2ef} shows that across all models, \gls{RS2EF} metrics are considerably worse than their \gls{S2EF} counterparts. Similar to \gls{OC22}, we see \gls{OOD} metrics to be significantly worse than \gls{ID}. Table \ref{tab:oc20-ris2re} shows a similar trend of \gls{RIS2RE} with worse performance. It is worth noting that total energy based metrics are a more challenging task than their referenced counter parts. A model tasked with predicting total energies is required to capture all subsurface, surface, and adsorbate interactions accurately. In the case of \gls{OC20}'s adsorption reference, because a slab reference energy is subtracted off, models are ultimately focused on the energy associated with only the adsorbate-surface interface.

\begin{table}
\caption{A comparison of \gls{OC20} performance on \gls{S2EF} and \gls{RS2EF}. Across all models and splits, \gls{RS2EF}, results in worse performance. Results are reported on the \gls{OC20} validation splits.}
\label{tab:oc20-rs2ef}
\resizebox{0.8\textwidth}{!}{%
\begin{tabular}{@{}llSSS@{}}
\toprule
Task & Model & {Energy MAE [eV] $\downarrow$} & {Force MAE [eV/\AA] $\downarrow$} & {Force Cosine $\uparrow$} \\ \midrule \multicolumn{5}{c}{ID} \\ \multirow{3}{*}{\gls{S2EF}} & SchNet & 0.4468 & 0.0493 & 0.3185 \\
 & DimeNet++ & 0.4545 & 0.0443 & 0.3632 \\
 & GemNet-dT & 0.2416 & 0.0227 & 0.6133 \\ \midrule
\multirow{3}{*}{\gls{RS2EF}} & SchNet & 3.7370 & 0.0473 & 0.3426 \\
 & DimeNet++ & 3.0430 & 0.0316 & 0.5145 \\
 & GemNet-dT & 0.4657 & 0.02465 & 0.5859 \\ \midrule\midrule
 \multicolumn{5}{c}{OOD Ads} \\ 
\multirow{3}{*}{\gls{S2EF}} & SchNet & 0.4973 & 0.0574 & 0.2862 \\
 & DimeNet++ & 0.5093 & 0.0508 & 0.3401 \\
 & GemNet-dT & 0.2465 & 0.0254 & 0.6052 \\ \midrule
\multirow{3}{*}{\gls{RS2EF}} & SchNet & 3.7560 & 0.0534 & 0.3181 \\
 & DimeNet++ & 3.0520 & 0.0351 & 0.5028 \\
 & GemNet-dT & 0.4732 & 0.0279 & 0.5859 \\ \midrule\midrule
 \multicolumn{5}{c}{OOD Cat} \\
\multirow{3}{*}{\gls{S2EF}} & SchNet & 0.5453 & 0.0520 & 0.2973 \\
 & DimeNet++ & 0.5184 & 0.0445 & 0.3512 \\
 & GemNet-dT & 0.3572 & 0.0269 & 0.5610 \\ \midrule
\multirow{3}{*}{\gls{RS2EF}} & SchNet & 3.8530 & 0.0492 & 0.3119 \\
 & DimeNet++ & 3.1590 & 0.0341 & 0.4715 \\
 & GemNet-dT & 1.0330 & 0.0306 & 0.5260 \\ \midrule\midrule
 \multicolumn{5}{c}{OOD Both} \\
\multirow{3}{*}{\gls{S2EF}} & SchNet & 0.7047 & 0.0685 & 0.2854 \\
 & DimeNet++ & 0.6753 & 0.0589 & 0.3556 \\
 & GemNet-dT & 0.4149 & 0.0335 & 0.5963 \\ \midrule
\multirow{3}{*}{\gls{RS2EF}} & SchNet & 4.7700 & 0.0639 & 0.3132 \\
 & DimeNet++ & 3.9460 & 0.0428 & 0.5000 \\
 & GemNet-dT & 1.2630 & 0.0386 & 0.5599 \\ \bottomrule
\end{tabular}%
}
\end{table}
\begin{table}
\caption{A comparison of \gls{OC20} performance on \gls{IS2RE} and \gls{RIS2RE}. Across all models and splits, \gls{RIS2RE} results in worse performance.}
\label{tab:oc20-ris2re}
\resizebox{0.7\textwidth}{!}{%
\begin{tabular}{@{}llSS[round-precision=2]@{}}
\toprule
Task & Model & {Energy MAE [eV] $\downarrow$} & {EwT [\%] $\uparrow$} \\ \midrule \multicolumn{4}{c}{ID} \\ \multirow{3}{*}{\gls{IS2RE}}
 & SchNet    & 0.6372 &	2.96 \\
 & DimeNet++ & 0.5605 &	4.26 \\
 & GemNet-dT & 0.5264 &	4.65 \\ \midrule
\multirow{3}{*}{\gls{RIS2RE}} 
 & SchNet & 1.6980 &	0.87 \\
 & DimeNet++ & 1.3780 &	1.35 \\
 & GemNet-dT & 1.1960 &	1.41 \\ \midrule\midrule
 \multicolumn{4}{c}{OOD Ads} \\ 
\multirow{3}{*}{\gls{IS2RE}}
 & SchNet & 0.7342 &	2.33 \\
 & DimeNet++ & 0.7252 &	2.06 \\
 & GemNet-dT & 0.7053 &	2.21 \\
 \midrule
\multirow{3}{*}{\gls{RIS2RE}} 
 & SchNet    & 1.6830 &	1.05 \\
 & DimeNet++ & 1.4260 &	1.19 \\
 & GemNet-dT & 1.2180 &	1.33 \\ \midrule\midrule
 \multicolumn{4}{c}{OOD Cat} \\
\multirow{3}{*}{\gls{IS2RE}}
 & SchNet    & 0.6611 & 2.95 \\
 & DimeNet++ & 0.5750 &	4.10 \\
 & GemNet-dT & 0.5326 &	4.59 \\
\midrule
\multirow{3}{*}{\gls{RIS2RE}} 
 & SchNet    & 2.3590 &	0.64 \\
 & DimeNet++ & 1.7660 & 0.93 \\
 & GemNet-dT & 2.1020 &	0.77 \\ \midrule\midrule
 \multicolumn{4}{c}{OOD Both} \\
\multirow{3}{*}{\gls{IS2RE}}
 & SchNet    & 0.7035 & 2.22 \\
 & DimeNet++ & 0.6613 &	2.42 \\
 & GemNet-dT & 0.6433 &	2.31 \\ \midrule
\multirow{3}{*}{\gls{RIS2RE}} 
 & SchNet    & 2.5760 &	0.59 \\
 & DimeNet++ & 1.9940 &	0.80 \\
 & GemNet-dT & 2.3310 &	0.72 \\ \bottomrule
\end{tabular}%
}
\end{table}
\clearpage

\section{Alternative reference scheme}
Whereas \gls{OC20} references systems to correspond to an adsorption energy, \gls{OC22} is only concerned with making total energy predictions. In the context of model training, an adsorption energy reference modifies the target energy distribution of the dataset. Normalization schemes have been known to aid in accelerating and improving model training, particularly for deep neural networks \cite{huang2020normalization}. We present a ``linear referencing" approach as a normalization scheme for \gls{OC22}. First, we fit a linear regression model to learn per-atom energies, i.e.

\begin{align*}
    \bm{K} &= \begin{bmatrix}
K^1_H & K^1_{He} & K^1_{Li} & \dots & K^1_{Fm}\\
K^2_H & K^2_{He} & K^2_{Li} & \dots & K^2_{Fm}\\
\vdots & \vdots & \vdots & \vdots & \vdots &\\
K^N_H & K^N_{He} & K^N_{Li} & \dots & K^N_{Fm}\\
\end{bmatrix}, \bm{P} = \begin{bmatrix}
E_H\\
E_{He}\\
\vdots\\
E_{Fm}
\end{bmatrix}
\end{align*}

\begin{equation}
\bm{K}\bm{P}
= 
\begin{bmatrix}
E^{DFT}_1\\
E^{DFT}_2\\
\vdots\\
E^{DFT}_{N}
\end{bmatrix}
\end{equation}

Where $K^{i}_{j}$ corresponds to the count of element $j$ in system $i$, $E_j$ the per-element energy being fit, and $E^{DFT}_i$ the ground truth DFT  total energy. Once fitted, energy targets used for training, $E^{ML}_i$, are referenced as follows:

\begin{equation}
    E^{ML}_{i} = E^{DFT}_{i} - \bm{K^{i}}\bm{P}
\end{equation}

Where $\bm{K^{i}}$ are the element counts for system $i$ and $\bm{P}$ is the set of fitted per-element energy coefficients. 
Table \ref{tab:linref} compares model performance on \gls{RS2EF} with and without the proposed linear reference. Across all models, we see a 4.5\%, 49.9\%, and 24.5\% improvement in \gls{ID} energy metrics for GemNet-OC\cite{gasteiger2022graph}, GemNet-dT\cite{gasteiger2021gemnet}, and SpinConv\cite{shuaibi2021rotation}, respectively. With the execption of GemNet-OC, all models see similar energy improvements for the \gls{OOD} split. As expected, force metrics across all models see little change. This referencing was omitted from the main paper as to encourage other strategies to energy normalization, particularly for large, diverse datasets like \gls{OC22} and \gls{OC20}. 

\begin{table}
\caption{\gls{OC22} \gls{RS2EF} test results for several top performing baseline \gls{GNNs}, with and without a linear referencing scheme. A linear reference serves as an energy normalization strategy, aiding in overall energy performance across all models.}
\label{tab:linref}
\resizebox{\columnwidth}{!}{%
\begin{tabular}{@{}llSSSSSS@{}}
\toprule
 &  & \multicolumn{3}{c}{ID} & \multicolumn{3}{c}{OOD} \\ \midrule
\multirow{2}{*}{Reference} & \multirow{2}{*}{Model} & {Energy} & {Force} & {Force} & {Energy} & {Force} & {Force} \\
 & & {MAE [eV] $\downarrow$} & {MAE [eV/\AA] $\downarrow$} & {Cosine $\uparrow$} & {MAE [eV] $\downarrow$} & {MAE [eV/\AA] $\downarrow$} & {Cosine $\uparrow$}\\\midrule
\multirow{3}{*}{None} & SpinConv & 0.8364 &	0.0377 & \multicolumn{1}{c|}{0.5905} & 1.9440 &	0.0631 &	0.4118 \\
 & GemNet-dT & 0.9385 &	0.0316  & \multicolumn{1}{c|}{0.6647} & 1.2713 &	0.0405 &	0.5303 \\
 & GemNet-OC & 0.3742 &	0.0294 & \multicolumn{1}{c|}{0.6913} & 0.8287 &	0.0396 &	0.5500 \\ \midrule
\multirow{3}{*}{Linear} & SpinConv & 0.6314 &	0.0374  & \multicolumn{1}{c|}{0.5988} & 1.4240 &	0.0686 &	0.4044 \\
 & GemNet-dT & 0.4699 &	0.0315 & \multicolumn{1}{c|}{0.6723} & 1.0912 &	0.0423 &	0.5278 \\
 & GemNet-OC & 0.3572 &	0.0297 & \multicolumn{1}{c|}{0.6922} & 1.0569 &	0.0396 &	0.5515 \\ \bottomrule
\end{tabular}%
}
\end{table}
\clearpage

\section{Use of total energy models to predict adsorption energies}
Total \gls{DFT} energy predictions, although more general than adsorption energies, are not physically meaningful without a proper reference. One benefit to total energy prediction, however, is that there are many possible catalysis and materials relevant properties that can be computed, only one of which is the adsorption energy. Here we elaborate on adsorption energy predictions via a direct route, as was done in \gls{OC20}, and via total energy predictions, as is the case for \gls{OC22}. The first approach directly predicts adsorption energy as $ E_{ad} = \hat{E}_{ad}$. Where models are trained on adsorption energy targets ($E_{ad} = E_{sys} - E_{slab} - E_{gas}$) that make use of DFT for the adsorbate+slab system, the clean slab, and the gas-phase adsorbate reference. For the total energy models we consider two approaches for calculating the adsorption energy \emph{Mixed-ML} (Main Equation 4) and \emph{Full-ML} (Main Equation 5).  

\begin{figure}
    \centering
    \includegraphics[width=\columnwidth]{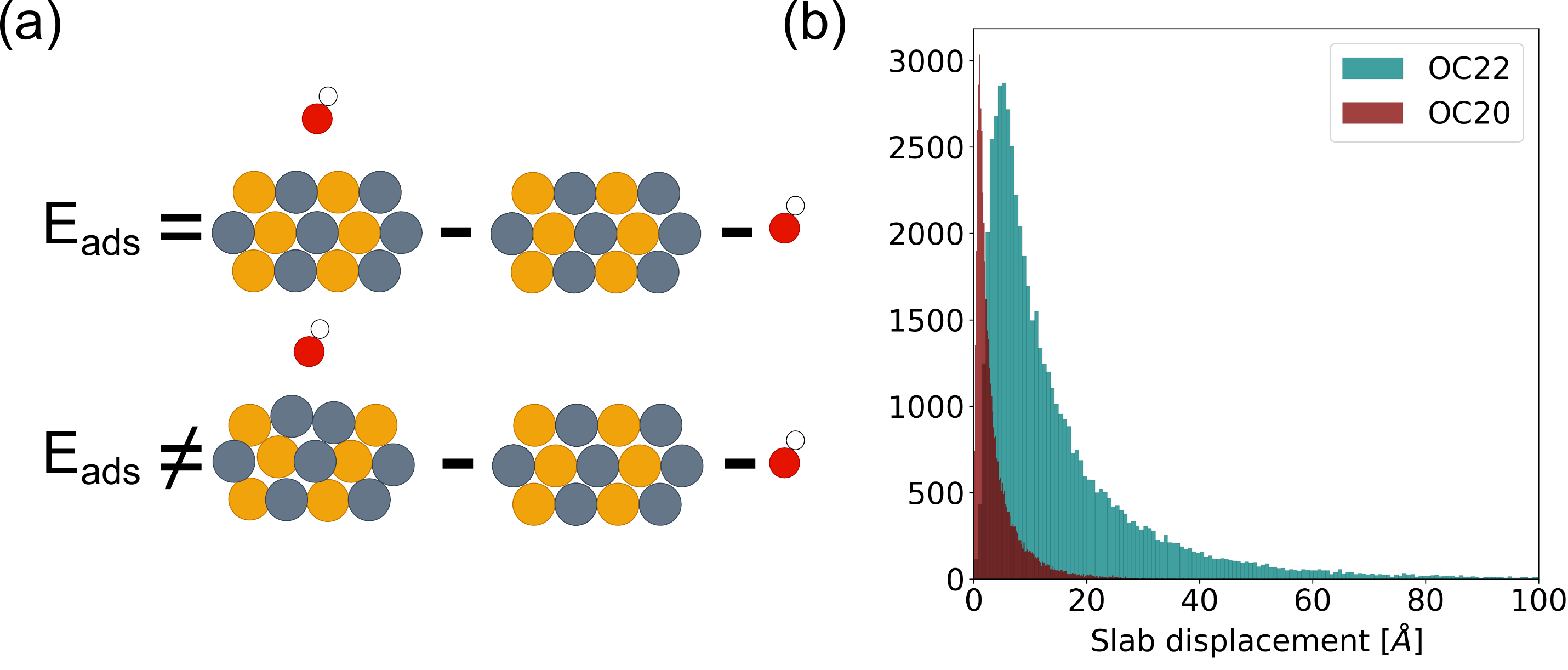}
    \caption{\textbf{(a)} A correct adsorption calculation assumes the relaxed adsorbate+slab and relaxed clean slab reference are consistent. \textbf{(b)} A histogram of cumulative slab displacement between the relaxed adsorbate+slab and relaxed clean slab. \gls{OC22} systems observed a significant amount of movement compared to \gls{OC20}, a consequence of all \gls{OC22} atoms being unconstrained and slabs not being optimized before adsorbate placement.}
    \label{fig:adsref}
\end{figure}

The motivation for the change in referencing is two-fold: it enables adsorption energy predictions that span different surface coverages that are particularly important in oxide catalysts and it opens up the possibility of new property predictions such as surface energies. In order to maximize the properties accessible in the dataset we allowed all the slab atoms to relax. However, this adds the complication of potentially having possible inconsistent slab references. We note that to make an accurate adsorption energy calculation, the corresponding relaxed slab reference needs to be identical or similar to that of the relaxed adsorbate+slab (Figure \ref{fig:adsref}a). Since \gls{OC22} created adsorbate+slabs from unrelaxed slabs and relaxed corresponding pairs in parallel, it is possible that the slab is no longer a consistent reference. Figure \ref{fig:adsref}b illustrates the cumulative slab atom displacement between the relaxed adsorbate+slab and relaxed clean slab for both \gls{OC20} and \gls{OC22}.

In order to validate our Mixed-ML and Full-ML adsorption energy predictions, we subsampled the OC22 validation set for $\sim$ 700 systems and reran DFT to get the appropriate adsorption energy reference. In order to get a meaningful sample size we had to sample some of the clean slabs from the training set. The DFT settings were identical to the original calculations with the exception that subsurface atoms were fixed (selective dynamics), and relaxations were done in series not parallel. So, the slab was first relaxed then the adsorbate was placed on the relaxed slab and then the slab+adsorbate system was relaxed. In practice, one would use the same serial procedure for Full-ML predictions, however, in our adsorption energy validation study we used the DFT initial structures to make ML predictions. This was done to ensure the input structures to DFT and the ML model were identical and, at least in theory, could relax to the same energy minimum. As a result, the slab in the slab+adsorbate system was a DFT relaxed slab. Overall, this makes the ML prediction problem easier, albeit necessary for a true comparison, and should be acknowledged when interpreting the results in Main Table 8. 

\clearpage
\section{Atomic drift with unconstrained relaxation}
\gls{OC22} systems observed a significant amount of cumulative movement compared to \gls{OC20} with an average drift of 14.83 Å and 3.80 Å in atomic positions for \gls{OC22} and \gls{OC20} respectively. This is a consequence of all \gls{OC22} atoms being unconstrained and slabs not being optimized before adsorbate placement. Despite the significantly larger cumulative displacement however, the per-atom slab displacement for \gls{OC22} (0.220 Å) is consistent with that of \gls{OC20} (0.231 Å). In a slab, the majority of atoms will consist of bulk-like atoms which will contribute more to the cumulative drift exhibited during full relaxation. The small drift in \gls{OC22} indicates that the over all positions of these bulk-like atoms however remain consistent with a bulk structure and by extension, the bulk-like properties of the center of the slab, within 0.220 Å. This result indicates that for the majority of slabs, full relaxation will not lead to significant reconstruction.

\clearpage
\section{\gls{OC22} adsorbate+slab and slab only performance}
The \gls{OC22} dataset contains a combination of both adsorbate+slab (adsorbate on a slab) and slab systems. While evaluation metrics are averaged across all systems, it may be useful to explore the performance of a particular subset. We trained several models on subsets in isolation and report results in Table \ref{tab:subsets}. Adsorbate+slab performance does considerably better than slabs, a possible consequence of dataset size and the nature of adsorbate+slab relaxations sampling a larger configurational space (e.g. slab relaxations often only require a few dozen \gls{DFT} calculations). Best adsorbate+slab and slab performance is achieved when training on both subsplits, suggesting adsorbate+slabs are useful to improving slab performance, and vice-versa.
\begin{table}
\caption{\gls{RS2EF} results on adsorbate+slab (adslabs) and slab subsets of the \gls{OC22} test splits. Models were trained and evaluated on only that subset. GemNet-OC* corresponds to the baseline model trained on all of \gls{OC22} but evaluated on the subsets in isolation.}
\label{tab:subsets}
\resizebox{0.8\textwidth}{!}{%
\begin{tabular}{@{}llSSSSSS@{}}
\toprule
\multirow{2}{*}{Subset} & \multirow{2}{*}{Model} & \multicolumn{2}{c}{Energy MAE [eV] $\downarrow$} & \multicolumn{2}{c}{Force MAE [eV/\AA] $\downarrow$} & \multicolumn{2}{c}{Force Cosine $\uparrow$} \\ \cmidrule(l){3-8} 
 &  & {ID} & {OOD} & {ID} & {OOD} & {ID} & {OOD} \\ \midrule
\multirow{4}{*}{Adslabs} 
 & SpinConv &1.0226 &	1.9972 & 0.0353 &	0.0632 &  0.5921  &	0.4085 \\
 & GemNet-dT & 0.8884 &	1.4014&  0.0302 &	0.0395  & 0.6595 &	0.5273  \\
 & GemNet-OC &  0.4406 &	0.8669& 0.0287 &	0.0365  & 0.6734 &	0.5480  \\
 & GemNet-OC * & 0.3488 &	0.8108 & 0.0272 &	0.0370 & 0.6950 &	0.5534\\ \midrule
\multirow{4}{*}{Slabs} 
 & SpinConv & 1.5422 &	2.2083  &  0.0741 &	0.0918 &  0.4972 &	0.3821  \\
 & GemNet-dT & 1.6132 &	2.0378  & 0.0567 &	0.0617 &  0.4910 &	0.4182  \\
 & GemNet-OC & 1.1230 &	1.6587  & 0.0523 &	0.0547  & 0.5079 &	0.4351 \\
 & GemNet-OC* & 0.4744 &	0.9024 & 0.0374 &	0.0499 & 0.6776 &	0.5365 \\ \bottomrule
\end{tabular}%
}
\end{table}

\section{Training and hyperparameters}
Baseline models used hyperparameters originally from \gls{OC20} or included a light sweep over some of the training settings - learning rate, optimizer, scheduler. For experiments within a particular task, model architectures were fixed. All model hyperparameters will be accessible at \url{https://github.com/Open-Catalyst-Project/ocp/tree/main/configs/oc22}. \gls{RS2EF} models trained only on \gls{OC22} used an atomwise loss function\cite{batzner20223, musaelian2022learning} - weighing energy+forces in the loss function by $1:N^2_{atoms}$. We found this to improve force metrics. A stepwise learning rate scheduler was also used for these experiments, decaying the learning rate at 2,3,4,5,6 epochs. \gls{RS2EF} joint training jobs used the original \gls{OC20} loss function and used a reduce-on-plateau learning rate scheduler. \gls{RS2EF} fine-tuning experiments all used the original \gls{OC20} loss function as we noticed an atomwise loss function overfit very quickly on forces. The \gls{OC22}-only experiments of the main paper used the \gls{OC20} loss function to allow for direct comparisons with the fine-tuning experiments. All joint training experiments involving \gls{OC20} used  DFT total energies instead of adsorption energies. Unlike \gls{OC20}, no energy normalization was done for training as we saw it to hurt performance across the board.

Models were trained using anywhere from 4-64 GPUs on 32Gb NVIDIA Volta cards. Learning rates for all fine-tuning experiments were reduced by 5-10x as compared to their base counterparts to ensure stable training. All models were optimized using AMSGrad. We provide the hyperparameters for our best performing \gls{RS2EF} model variant, GemNet-OC, for the joint and fine-tuning training strategies in Table \ref{tab:goc}.

\begin{table}
\caption{Model hyperparameters for the top performing GemNet-OC joint and fine-tuning experiments.}
\label{tab:goc}
\resizebox{0.8\textwidth}{!}{%
\begin{tabular}{lc@{\hspace{0.2cm}}c@{\hspace{0.2cm}}c}
Hyperparameters                 &       &   \gls{OC20}+\gls{OC22}     &   \gls{OC20}$\rightarrow$\gls{OC22}     \\
\hline
No. spherical basis     & & 7  & 7 \\
No. radial basis      & & 128   & 128\\
No. blocks            & & 4          & 4 \\
Atom embedding size                       & & 256         & 256\\
Edge embedding size                    & & 512         & 512\\\\
Triplet edge embedding input size                   & & 64        & 64\\
Triplet edge embedding output size                      & & 64          & 64 \\
Quadruplet edge embedding input size         & & 32           & 32\\
Quadruplet edge embedding output size     & & 32        & 32\\
Atom interaction embedding input size               & & 64             & 64\\
Atom interaction embedding output size               & & 64             & 64\\
Radial basis embedding size   & & 16         & 16\\
Circular basis embedding size   & & 16         & 16\\
Spherical basis embedding size   & & 32         & 32\\\\
No. residual blocks before skip connection & & 2         & 2\\
No. residual blocks after skip connection & & 2         & 2\\
No. residual blocks after concatenation & & 1         & 1\\
No. residual blocks in atom embedding blocks & & 3         & 3\\
No. atom embedding output layers & & 3         & 3\\\\
Cutoff  & & 12.0         & 12.0\\
Quadruplet cutoff  & & 12.0         & 12.0\\
Atom edge interaction cutoff  & & 12.0         & 12.0\\
Atom interaction cutoff  & & 12.0         & 12.0\\
Max interaction neighbors  & & 30         & 30\\
Max quadruplet interaction neighbors  & & 8         & 8\\
Max atom edge interaction neighbors  & & 20         & 20\\
Max atom interaction neighbors  & & 1000         & 1000\\\\
Radial basis function & & Gaussian         & Gaussian\\
Circular basis function & & Spherical harmonics   & Spherical harmonics\\
Spherical basis function & & Legendre Outer         & Legendre Outer\\
Quadruplet interaction & & True      & True\\
Atom edge interaction & & True         & True\\
Edge atom interaction & & True         & True\\
Atom interaction & & True       & True\\
Direct forces & & True         & True\\\\
Activation & & Silu        & Silu\\
Optimizer & & AdamW & AdamW\\
EMA decay & & 0.999 & 0.999 \\
Gradient clip norm threshold & & 10 & 10\\
Learning rate & & 0.0005 & 0.0001\\
Scheduler & & ReduceLROnPlateau & StepwiseLRDecay\\
LR Milestones & & N/A & epochs 2-10, 0.5 after \\
Force loss function & & AtomwiseL2 & L2 \\
Energy loss function & & MAE & MAE \\
Force coefficient & & 1 & 100\\
Energy coefficient & & 1 & 1\\
\end{tabular}
}
\end{table}

\clearpage

\section{\gls{RS2EF}, \gls{RIS2RE}, \gls{IS2RS} validation results}
Full validation results are shown in Tables \ref{tab:rs2ef-val}, \ref{tab:ris2re-val}, \ref{tab:is2rs-val} for \gls{RS2EF}, \gls{RIS2RE}, and \gls{IS2RS}, respectively.

\begin{table*}
\caption{Predicting total energy and force from a structure (\gls{RS2EF}). Results are shared for the default, joint training, and fine-tuning training strategies. Experiments are evaluated on the validation set.}
\label{tab:rs2ef-val}
\resizebox{\textwidth}{!}{%
\begin{tabular}{@{}clllllllSSlSSlSSlS[round-precision=2]S[round-precision=2]l@{}}
& \multicolumn{19}{c}{\gls{RS2EF} Validation}  \\
\cmidrule(r){1-19}
\mc{3}{l}{\mr{2}{*}{Training}} &
   &
  \mc{3}{l}{\mr{2}{*}{Model}} &
   &
  \mc{2}{c}{Energy MAE [eV] $\downarrow$} &
  \mc{1}{c}{} &
  \mc{2}{c}{Force MAE [eV/\AA] $\downarrow$} &
  \mc{1}{c}{} &
  \mc{2}{c}{Force Cosine $\uparrow$} &
  \mc{1}{c}{} &
  \mc{2}{c}{EFwT [\%] $\uparrow$} &
   \\ \cmidrule(lr){8-10} \cmidrule(lr){11-13} \cmidrule(lr){14-16} \cmidrule(lr){17-19}
\mc{3}{l}{} &
   &
  \mc{3}{c}{} &
   &
  {ID} &
  {OOD} &
  \mc{1}{c}{} &
  {ID} &
  {OOD} &
  \mc{1}{c}{} &
  {ID} &
  {OOD} &
  \mc{1}{c}{} &
  {ID} &
  {OOD} &
   \\ \cmidrule(r){1-19}
\mc{3}{l}{\mr{9}{*}{OC22-only}}      &  &
\mc{3}{l}{Median Baseline} &  & 169.7327	& 164.3158 &  & 0.0756 &	0.0737 &  & 0.0023 &	0.0018 &  & 0.00 &	0.00 &  \\
\mc{3}{c}{}                           &  &
\mc{3}{l}{SchNet~\cite{schutt2017schnet}} &  & 8.0790 &	12.5600 &  & 0.0599 &	0.0917 &  & 0.3591 &	0.2326 &  & 0.00 &	0.00 &  \\
\mc{3}{c}{}                           &  &
\mc{3}{l}{DimeNet$++$~\cite{klicpera2020directional,klicpera2020fast}} &  & 2.3550 &	2.8410 &  & 0.0433 &	0.0606 &  & 0.5989 &	0.4588 &  & 0.00 &	0.00 &  \\
\mc{3}{c}{}                           &  &
\mc{3}{l}{ForceNet~\cite{hu2021forcenet}} &  & {-} & {-} &  & 0.0569 &	0.0658 &  & 0.3447 &	0.3022 &  & 0.00 &	0.00 &  \\
\mc{3}{c}{}                           &  &
\mc{3}{l}{SpinConv~\cite{shuaibi2021rotation}} &  & 0.9847 &	2.2620 &  & 0.0384 &	0.0779 &  & 0.5843 &	0.4137 &  & 0.00 & 0.00 &  \\
\mc{3}{c}{}                           &  &
\mc{3}{l}{PaiNN~\cite{schutt2021equivariant}} &  & 1.1250 &	2.9480 &  & 0.0455 &	0.0617 &  & 0.4783 &	0.3643 &  & 0.00 &	0.00 &  \\
\mc{3}{c}{}                           &  &
\mc{3}{l}{GemNet-dT~\cite{gasteiger2021gemnet}} &  & 1.1060 &	1.8430 &  & 0.0322 &	0.0408 &  & 0.6565 &	0.5597 &  & 0.01 &	0.00 &  \\
\mc{3}{c}{}                           &  &
\mc{3}{l}{GemNet-OC~\cite{gasteiger2022graph}} &  & 0.5445 &	1.0110 &  & 0.0303 &	0.0403 &  & 0.6831 &	0.5795 &  & 0.03 &	0.00 &  \\
\midrule
\midrule
\mc{3}{l}{\mr{3}{*}{OC20-2M + OC22}}      &  &
\mc{3}{l}{PaiNN\cite{schutt2021equivariant}} &  & 0.5717 &	1.5760 &  & 0.0482 &	0.0688 &  & 0.4601 &	0.3372 &  & 0.02 &	0.00  \\
\mc{3}{c}{}                           &  &
\mc{3}{l}{SpinConv\cite{shuaibi2021rotation}} &  & 1.1250 &	2.1470 &  & 0.0368	& 0.0648 &  & 0.6145 &	0.4594 &  & 0.00 &	0.00 & \\
\mc{3}{c}{}                           &  &
\mc{3}{l}{GemNet-OC~\cite{gasteiger2022graph}} &  & 0.6047 &	1.0930 &  & 0.0295 &	0.0382 &  & 0.6846 &	0.5886 &  & 0.04 &	0.01 &  \\
\midrule
\mc{3}{l}{\mr{3}{*}{OC20-20M + OC22}}      &  &
\mc{3}{l}{PaiNN\cite{schutt2021equivariant}} &  & 0.5422 &	1.3210 &  & 0.0466 &	0.0640 &  & 0.4720 &	0.3593 &  & 0.02 &	0.00  \\
\mc{3}{c}{}                           &  &
\mc{3}{l}{SpinConv\cite{shuaibi2021rotation}} &  & 1.1370 &	2.2390 &  & 0.0365 &	0.0601 &  & 0.5942 &	0.4654 &  & 0.00 &	0.00 &  \\
\mc{3}{c}{}                           &  &
\mc{3}{l}{GemNet-OC~\cite{gasteiger2022graph}} &  & 0.5224 &	1.0950 &  & 0.0275 &	0.0362 &  & 0.7128 &	0.6145 &  & 0.09 &	0.02 &  \\
\midrule
\mc{3}{l}{\mr{2}{*}{OC20-All + OC22}}      &  &
\mc{3}{l}{SpinConv\cite{shuaibi2021rotation}} &  & 1.4880 &	2.3130 &  & 0.0405 &	0.0546 &  & 0.5226 &	0.4452 &  & 0.00 &	0.00 &  \\
\mc{3}{c}{}                           &  &
\mc{3}{l}{GemNet-OC~\cite{gasteiger2022graph}} &  & 0.4636 &	0.8585 &  & 0.0271 &	0.0339 &  & 0.6980 &	0.6167 &  & 0.10 &	0.01 &  \\
\midrule\midrule
\mc{3}{l}{\mr{4}{*}{OC20$\rightarrow$OC22}}      &  &
\mc{3}{l}{SpinConv\cite{shuaibi2021rotation}} &  & 1.3080 &	2.5690 &  & 0.0364 &	0.0569 &  &  0.5959	& 0.4649 &  & 0.00 &	0.00 \\
\mc{3}{c}{}                           &  &
\mc{3}{l}{GemNet-dT~\cite{gasteiger2021gemnet}} &  & 0.7189 &	1.0540 & & 0.0311 &	0.0412 && 0.6664 &	0.5728 && 0.02 &0.00 \\
\mc{3}{c}{}                           &  &
\mc{3}{l}{GemNet-OC~\cite{gasteiger2022graph}} &  & 0.3927 &	1.0400 &  & 0.0303 &	0.0400 &  & 0.6707 &	0.5689 &  & 0.12 &	0.00 &  \\
\mc{3}{c}{}                           &  &
\mc{3}{l}{GemNet-OC-Large~\cite{gasteiger2022graph}} &  & 0.4834 &	0.9635 & &0.0278 &	0.0389 && 0.7215 &	0.6110 && 0.03 &	0.02 \\
\bottomrule
\end{tabular}%
}
\end{table*}
\begin{table}
\caption{Predicting total relaxed energy from an initial structure (\gls{RIS2RE}). Results are shared for the default, joint training, and fine-tuning training strategies. Experiments are evaluated on the validation set.}
\label{tab:ris2re-val}
\resizebox{0.8\columnwidth}{!}{%
\begin{tabular}{@{}lllSSS[round-precision=2]S[round-precision=2]@{}}
\multicolumn{7}{c}{\gls{RIS2RE} Validation}  \\
\toprule
\multirow{2}{*}{Approach} & \multirow{2}{*}{Training} & \multirow{2}{*}{Model} & \multicolumn{2}{c}{Energy MAE [eV] $\downarrow$} & \multicolumn{2}{c}{EwT [\%] $\uparrow$} \\ \cmidrule(lr){4-5} \cmidrule(lr){6-7}
 &  &  & {ID} & {OOD} & {ID} & {OOD} \\ \midrule
\multirow{10}{*}{Direct} & \multirow{4}{*}{OC22-only}
 & Median Baseline & 183.9870 &	177.3488 & 0.00 &	0.00 \\
 &  & SchNet & 2.0190 &	5.2870 & 1.14 &	0.47 \\
 &  & DimeNet++ & 1.9920 &	4.3360 & 0.91 &	0.50 \\
 &  & PaiNN & 1.7702 &	4.3357 & 1.49 &	0.36 \\
 &  & GemNet-dT & 1.6900 &	4.5220 & 1.37 &	0.47 \\ \cmidrule(l){2-7} 
 & \multirow{4}{*}{OC20+OC22} 
 & SchNet & 3.0300 &	5.0760 & 0.65 &	0.43 \\
 &  & DimeNet++ & 1.9890 &	4.4500 & 0.91 &	0.61 \\
 &  & PaiNN & 1.7638 &	4.6901 & 1.33 &	0.36 \\
 &  & GemNet-dT & 2.5190 &	5.1500 & 0.61 &	0.40 \\ \cmidrule(l){2-7} 
 & \multirow{1}{*}{OC20$\rightarrow$OC22} 
 & GemNet-OC*& 1.2270 &	2.3600 & 4.08 &	1.08 \\
 \midrule
 \midrule
\multirow{8}{*}{Relaxation} & \multirow{3}{*}{OC22}
 & SpinConv & 1.7115 &	3.1760 & 1.37 &	0.58 \\
 &  & GemNet-dT & 1.8936 &	2.5747 & 1.07 &	0.83 \\
 &  & GemNet-OC & 1.3281 &	1.8834 & 2.06 &	1.29 \\ \cmidrule(l){2-7} 
 & \multirow{2}{*}{OC20+OC22} & SpinConv & 2.3131 &	3.4915 & 0.69 &	0.61 \\
 &  & GemNet-OC & 1.2472 &	2.0588 & 3.05 &	1.12 \\ \cmidrule(l){2-7} 
 & \multirow{3}{*}{OC20$\rightarrow$OC22} & SpinConv & 1.8780 &	3.4600 & 1.49 &	0.61 \\
 &  & GemNet-OC & 1.1730 &	1.9010 & 5.18 &	1.83 \\
  &  & GemNet-OC-Large & 1.2701 &	2.0399 & 1.30 &	1.22 \\
 \bottomrule
 \multicolumn{7}{l}{*GemNet-OC pretrained on \gls{OC20}+\gls{OC22} \gls{RS2EF}}  \\
\end{tabular}%
}
\end{table}
\begin{table}
\caption{Predicting relaxed structures from an initial structure \gls{IS2RS}. All models predicted relaxed structures through an iterative relaxation approach. The initial structure was used as a naive baseline (IS baseline). Experiments are evaluated on the validation set.}
\label{tab:is2rs-val}
\resizebox{0.5\textwidth}{!}{%
\begin{tabular}{@{}llS[round-precision=2]S[round-precision=2]@{}}
\multicolumn{4}{c}{\gls{IS2RS} Validation}  \\
\toprule
\multirow{2}{*}{Training} & \multirow{2}{*}{Model} & \multicolumn{2}{c}{ADwT [\%] $\uparrow$}\\ \cmidrule(lr){3-4} 
 &  & {ID} & {OOD} \\ \midrule
\multirow{4}{*}{OC22-only} 
 & IS baseline & 44.77 &	42.59  \\
 & SpinConv & 54.5270 &	40.4539    \\
 & GemNet-dT &  59.6791	& 51.2538  \\
 & GemNet-OC &  60.6946	& 52.8991  \\ \midrule
\multirow{2}{*}{OC20+OC22} 
 & SpinConv  &  55.7870  &	47.3090    \\
 & GemNet-OC &  60.9933  &	53.8461    \\ \midrule
\multirow{3}{*}{OC20$\rightarrow$OC22} 
 & SpinConv  & 56.6900  & 45.7800     \\
 & GemNet-OC &  58.0300 & 48.3300  \\
 & GemNet-OC-Large & 59.6940 & 51.6611    \\
 \bottomrule
\end{tabular}%
}
\end{table}
\clearpage

\section{Additional DFT settings}\label{section:more_vasps_settings}

All structure relaxations were performed using the Vienna ab initio simulation package (VASP)~\cite{Kresse1994, Kresse1996, Kresse1996a, vasp-license, kresse1999ultrasoft} with the projector augmented wave (PAW) approach. We modelled the exchange-correlation effects using the Perdew-Berke-Ernzerhof (PBE), generalized gradient approximation (GGA)~\cite{perdew1996generalized} which is generally accepted for modeling surface reactions on oxides\cite{Gonzalez2021, Heras-Domingo2019, VanDenBossche2017}. All calculations were performed with spin-polarization to account for the significant spin states in metal oxides. The external electrons were expanded in plane waves with kinetic energy cut-offs of 500 eV. The energies and atomic forces of all calculations were converged to within $1\times10^{-4}$ eV and 0.05 eV \AA$^{-1}$, respectively. We used $\Gamma$-centered \textit{k}-point meshes of $\frac{50}{a} \times \frac{50}{b} \times \frac{50}{c}$ and $\frac{30}{a} \times \frac{30}{b} \times 1$ for bulk and slab calculations, respectively, with non-integer values rounded up to the nearest integer. We used a Gaussian smearing algorithm for setting the partial occupancies of each orbital. We defaulted to a mixture of the blocked Davidson iteration\cite{Johnson2001} and the RMM-DIIS\cite{Wood1985, Pulay1980} scheme as the algorithm for electron minimization and withdrew to using only the blocked Davidson iteration for calculations containing Pb and In that failed to converge electronically. Ions were updated using the conjugated gradient algorithm. 

In this study, we placed adsorbates on one of the two surfaces of our slab which results in uneven charges between the two surfaces. This results in an nonphysical dipole moments that can lead to diverging total DFT energies. To account for this dipole moment, we introduced an electrostatic potential to the local potential of our adsorbate+slab.

As with OC20~\cite{oc20_perspective}, the value of $E_{gas}$ for each adsorbate was computed as a linear combination of \ce{N2}, \ce{H2O}, \ce{CO}, and \ce{H2} and is listed in Table~\ref{tab:per_atom_ref_e}. We note that we explicitly avoided using linear combinations involving \ce{O2} as the GGA functional is well known to overestimate the O-O bond strength.

\begin{table}
\caption{\label{tab:per_atom_ref_e} The per atom energy of individual adsorbate atoms used to calculate the gas phase reference energy for an adsorbate molecule.}
\begin{tabular}{c|c|c}
\toprule
Adsorbate atom & Energy (eV) & Linear combination\\
\hline
\ce{H} & -3.386 & $\frac{1}{2}$E(\ce{H2})\\
\ce{O} & -7.459 & E(\ce{H2O})-E(\ce{H2})\\
\ce{C} & -7.332 & E(\ce{CO})-E(\ce{O})\\
\ce{N} & -8.309 & $\frac{1}{2}$E(\ce{N2})\\
\hline
\end{tabular}
\end{table}

\section{Hubbard U corrections}
Materials with certain combinations of transition metals and oxygen are known to have strongly correlated electrons, i.e. the movement of electrons significantly influences the properties of other electrons. It is well known that the GGA functional is unable to properly account for these strong electron correlations leading to inaccurate calculations of thermodynamic and electronic properties. We account for this missing electron interaction by introducing the Hubbard U correction which uses a repulsive Coulombic force between the electrons. The strength of this repulsion stems from the ``U" value which is empirically fitted to experimental quantities such as the band gap or formation enthalpy. To properly account for the effects of strong electron correlation on the thermodynamic properties of our dataset, we adapted the Hubbard U values from the Materials Project which were fitted to correctly calculate the experimental enthalpy of formation\cite{Jain2011g} (see Table~\ref{tab:ggau} for the list of Hubbard U values).
\begin{table}
\caption{\label{tab:ggau} Hubbard U values for transition metals available on the Materials Project.}
\begin{tabular}{c|c}
\hline
& U (eV)\\
\hline
\hline
Co & 3.32 \\
\hline
Cr & 3.7 \\
\hline
Fe & 5.3 \\
\hline
Mn & 3.9 \\
\hline
Mo & 4.38 \\
\hline
Ni & 6.2 \\
\hline
V & 3.25 \\
\hline
W & 6.2 \\
\hline
\end{tabular}
\end{table}
\clearpage

\section{Chemical Diversity of Open Catalyst 2022}

\begin{figure}[H]
    \centering
    \includegraphics[width=0.6\textwidth]{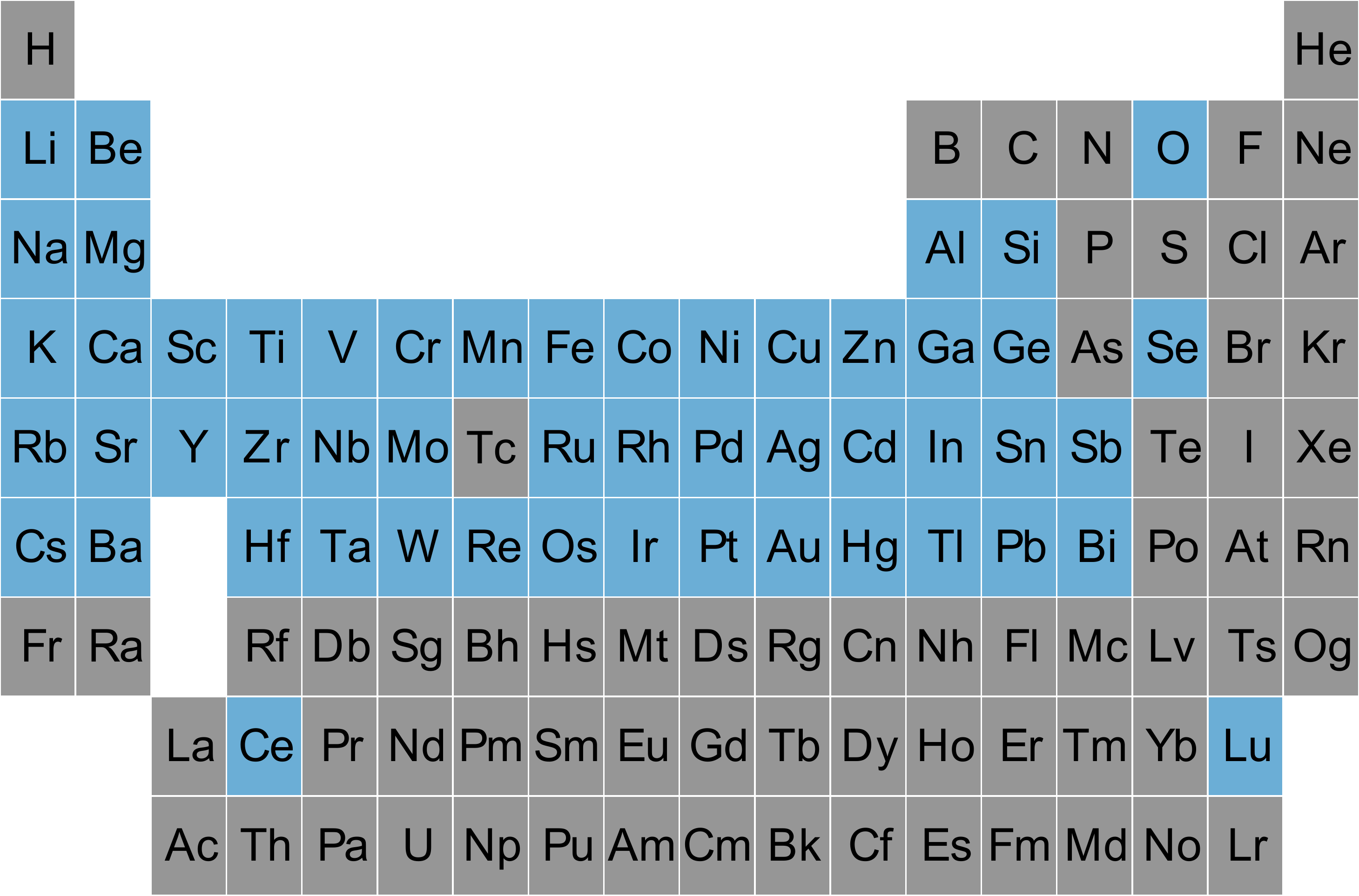}
    \caption{Periodic table showing the 51 elements considered in the OC22 dataset in blue. Elements that were not considered are show in grey. All slabs were constructed from bulk oxides composed of one (unary) or two (binary) of these metals.}
    \label{fig:periodic_table}
\end{figure}

\begin{figure}
    \centering
    \includegraphics[width=0.8\textwidth]{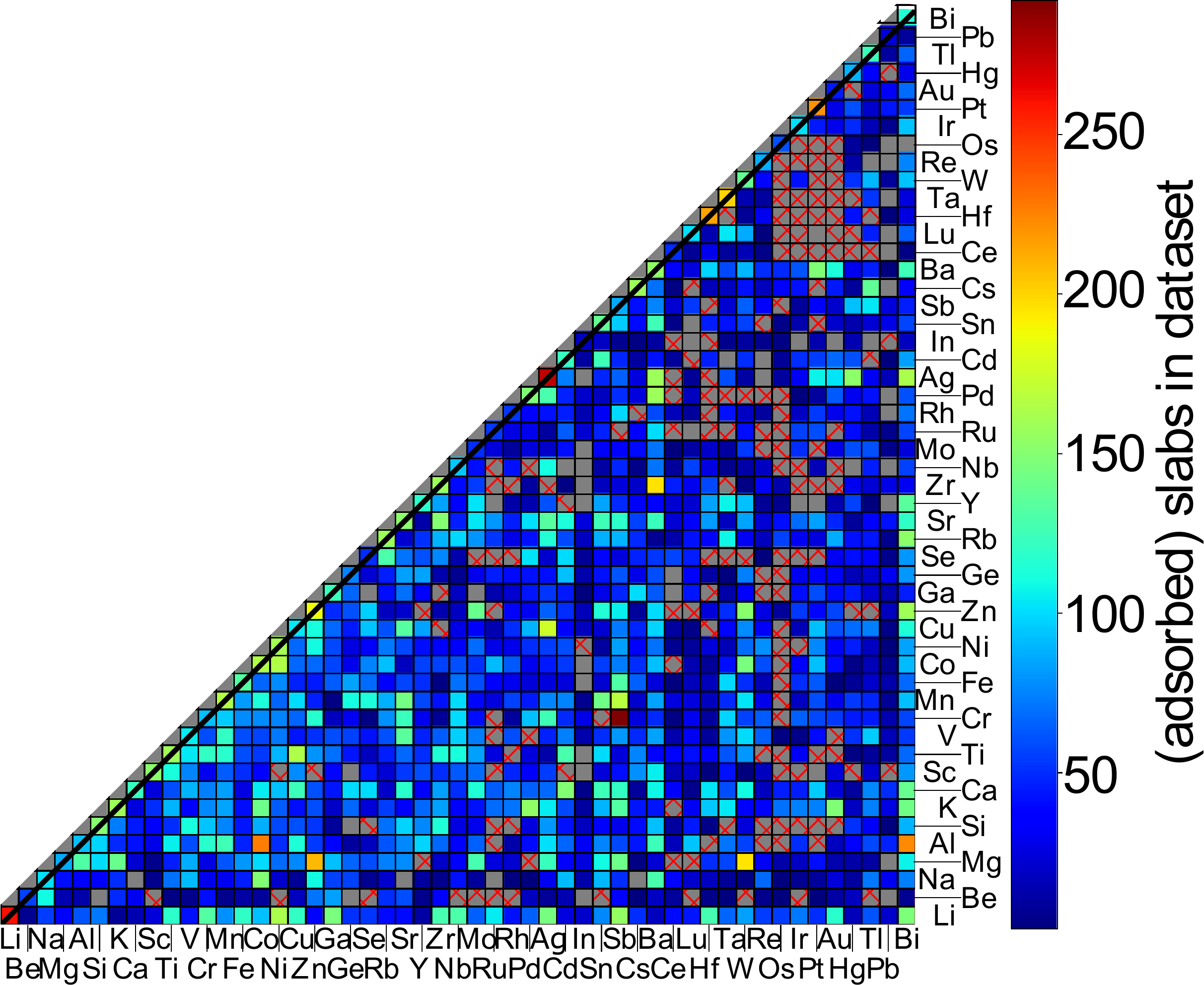}
    \caption{A 2D grid heat map indicating the number of slabs and adsorbate+slabs in the dataset containing specific pairs of metals of binary composition \ce{A_xB_yO_z}. Grid points on the diagonal correspond to unary compositions of \ce{A_xO_y}. Grey grids containing red hatches correspond to compositions that were not available in the Materials Project. Grey grids without hatches indicate compositions that were in our possible sample set of materials, but were not randomly sampled during the construction of the dataset.}
    \label{fig:compositional_sampling}
\end{figure}

The dataset of slabs in \gls{OC22} is constructed from a set of 51 elements shown in Figure~\ref{fig:periodic_table} resulting in $\binom{51}{1} = 51$ and $\binom{51}{2} = 1275$ possible unary and binary oxides respectively. We considered all transitions metals up to the 5d group with the exception of Tc due to its radioactivity (29 metals), all alkali and alkaline earth metals up to Fr (10), the lanthanides of Ce and Lu (2), and the p-block metals and metalloids: except Te and up to Po (11). We queried the materials project for materials with the top five lowest energy above hull and less than 150 atoms for all unary and binary oxides composed of these elements. This resulted in 4,728 bulk oxide structures considered. Figure~\ref{fig:compositional_sampling} provides a 2D grid heat map showing the frequency of chemical systems sampled in the OC22 dataset. Only 4,286 of the 4,728 bulk oxides were sampled in the dateset. Unary oxides are shown in the diagonal of the grid while all other blocks represent binary oxides. Not all chemical systems were sampled in the final dataset as some chemical systems did not exist in the Materials Project (red hatches) while other chemical systems had bulk oxide systems that were too large to create slabs of less than 150 atoms (grey blocks). We observe that for each chemical system considered, around 50 to 100 slabs and adsorbate+slabs were included in the final dataset which demonstrates the even distribution of chemical space sampled in the dataset. Slabs and adsorbate+slabs with Li-O, Sb-Cr-O and Ag-O were randomly over sampled with over 250 entries in the final dataset.

\pagebreak

\section{Gibbs free energy calculations}

The vibrational frequency and zero point energy of an atomistic system allows for the calculation of Gibbs free energy ($\Delta G$). This quantity allow us to determine Gibbs adsorption energy ($\Delta G_{ad}$) which is needed for constructing accurate reaction pathways, and creating microkinetic models to assess catalytic activity and selectivity. It also allows us to determine the overpotential, an important parameter used for screening viable catalysts in OER. Gibbs free energy can be determined with the following equation:

\begin{equation}
    \label{eq:gibbs_free_energy}
    \Delta G = E^{DFT}_{sys} + \Delta ZPE + \int_{0}^{T} C_p dT - T\Delta S
\end{equation}

where $E^{DFT}_{sys}$ is the total energy of the adsorbate+slab from DFT, $\Delta ZPE$ is the change in zero point energy, $\int_{0}^{T} C_p dT$ is the integral of the heat capacity ($C_p$) from a temperature of 0 K to T, T is the temperature and $\Delta S$ is the change in entropy. 

None of our datasets in the OC22 have consider the vibrational and entropic effects and as such, we did not construct any predictive models for $\Delta G$. However it is well known that, at specific atmospheric conditions, $\Delta G_{ad}$ can be obtained by shifting the DFT adsorption energy ($E_{ad}$) by a consistent correction value ($\Delta G_{corr}$) across all slabs with the same adsorbate\cite{gunasooriya2020analysis}. 

To determine $\Delta G_{corr}$, we randomly sampled 15-20 relaxed adsorbate+slabs per adsorbate (\ce{O^*}, \ce{OH^*}, and \ce{OOH^*}) from our dataset. Next, we performed subsequent  vibrational frequency calculations on the relaxed adsorbate+slabs by confining the entire slab and allowing only the adsorbate sites to relax. Post processing was performed using the the Vasp Gibbs\cite{Therrien} package to obtain $\Delta G$ at T = 298.15~K and P = 1~atm. We then computed Gibbs adsorption energy with the following:

\begin{equation}
    \label{eq:gibbs_adsorption_formula}
    \Delta G_{ad} = \Delta G_{sys} - E_{slab} - \Delta G_{gas}
\end{equation}

where $\Delta G_{sys}$ is the Gibbs free energy of the adsorbate+slab, $E_{slab}$ is the total DFT energy of the corresponding clean slab, and $\Delta G_{gas}$ represents the Gibbs formation energy of the adsorbate given by:
\begin{equation}
    \Delta G_{O_{(g)}} = \Delta G_{\ce{H2O_{(g)}}} - \Delta G_{\ce{H2_{(g)}}}
\end{equation}
\begin{equation}
    \Delta G_{OH_{(g)}} = \Delta G_{\ce{H2O_{(g)}}} - \frac{1}{2}\Delta G_{\ce{H2_{(g)}}}
\end{equation}
\begin{equation}
    \Delta G_{OOH_{(g)}} = 2\Delta G_{\ce{H2O_{(g)}}} - \frac{3}{2}\Delta G_{\ce{H2_{(g)}}}
\end{equation}

\begin{figure}
    \centering
    \includegraphics[width=0.5\textwidth]{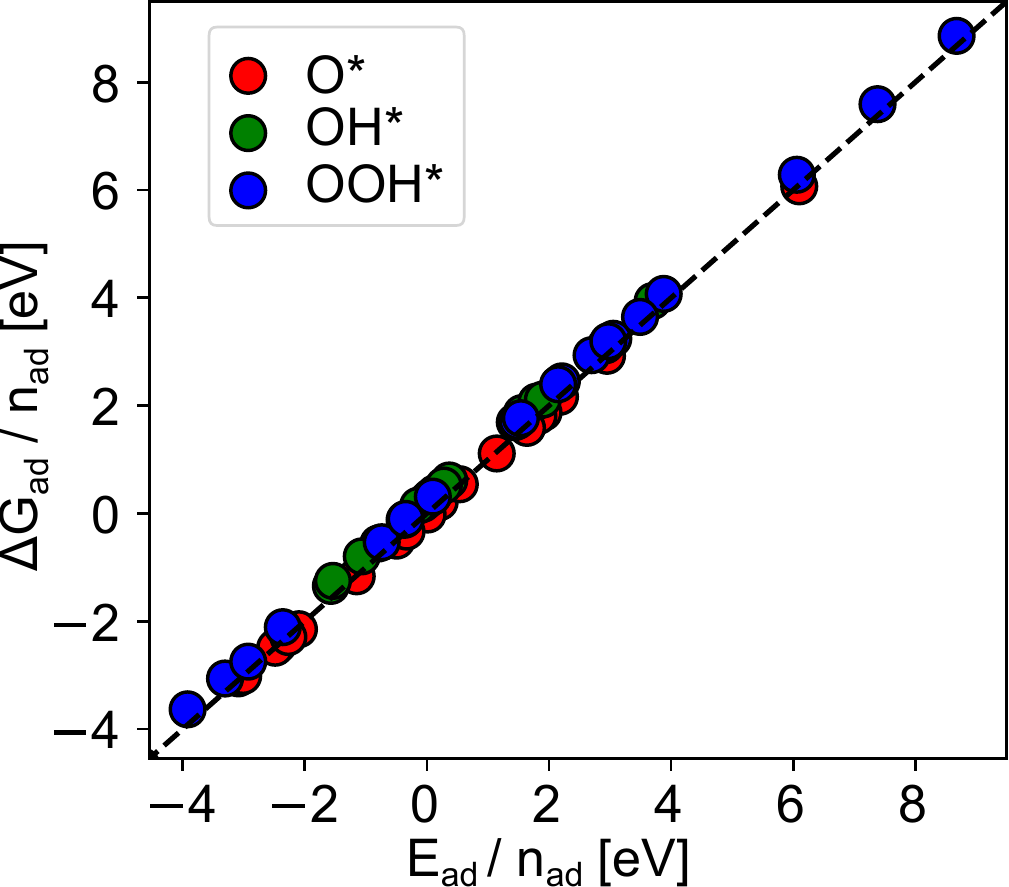}
    \caption{Randomly sampled DFT adsorption energies (x-axis) are plotted against their corresponding Gibbs adsorption energies (y-axis) for \ce{O^*}, \ce{OH^*}, and \ce{OOH^*}. The Gibbs adsorption energy was calculated using Equation~\ref{eq:gibbs_adsorption_formula}. A parity line is provided as the black dashed line. The fitted slope for all three adsorbates is 1. The intercept for each adsorbate are reported in Table~\ref{tab:gibbs_correction}.}
    \label{fig:gibbs_ads}
\end{figure}

Figure~\ref{fig:gibbs_ads} plots $\Delta G_{ad}$ against $E_{ad}$. The resulting fit between the two sets of data for \ce{O^*}, \ce{OH^*}, and \ce{OOH^*} showed a linear correlation with a slope close to unity and a non-zero intercept. This intercept represents $\Delta G_{corr}$ for adsorption and can be added to any value of $E_{ad}$ with the same adsorbate to obtain $\Delta G_{ad}$. $\Delta G_{corr}$ can similarly be determined for any adsorbate of our predicted quantities with minimal datapoints between $\Delta G_{ad}$ and $E_{ad}$. 

Table~\ref{tab:gibbs_correction} shows each of the average (across all datapoints in Figure~\ref{fig:gibbs_ads}) thermodynamic quantities calculated in this study to obtain $\Delta G_{corr}$ alongside the corresponding values obtained by~\citet{gunasooriya2020analysis}. All quantities in both studies for $\Delta G^{O^*}_{corr}$ and $\Delta G^{OH^*}_{corr}$ are consistent within 0.1 eV. However we observe a significant deviation in $\Delta G^{OOH^*}_{corr}$ with the value in this study underestimating the value obtained in the literature. This stems from a $25.89 J K^{-1} mol^{-1}$ difference in $\Delta S$ for \ce{H2O_{(g)}} which can change $\Delta G_{corr}$ by 0.06 eV. Although this may be due to the difference in functionals, it is worth noting that~\citet{East1997} has also reported a significantly lower entropy for water much closer to our values. Regardless, our results still demonstrate that the Gibbs adsorption energy can be obtained by calculating a systematic correction value that can be added to our predicted quantities without the need for additional intense computational efforts.

\begin{table}
\caption{\label{tab:gibbs_correction} Averaged thermodynamic quantities from Equation~\ref{eq:gibbs_free_energy} for each adsorbate+slab used to calculate $\Delta G_{corr}$. Values obtained from~\citet{gunasooriya2020analysis} are shown in parenthesis for comparison. The standard deviations for $\Delta G_{ads} - E_{ads}$ and intercepts in Figure~\ref{fig:gibbs_ads} are shown in the 7th and 8th column respectively. All values are in eV per adsorbate. Calculations were performed at a temperature of 298.15 and a pressure of 1 atm.}
\resizebox{1\columnwidth}{!}{%
\begin{tabular}{cccccccc}
\toprule
& $\bar{\Delta ZPE}$ & $\bar{\int_{0}^{T} C_p dT}$ & $\bar{-T \Delta S}$ & $\bar{\Delta G_{species}}$ & $\Delta G_{corr}$ & STDEV & intercept \\
\hline
\ce{H2O_{(g)}} & 0.52 (0.56) & 0.10 (0.10) & -0.60 (-0.68) & 0.02 (-0.01) & & & \\
\ce{H2_{(g)}} & 0.28 (0.27) & 0.09 (0.09) & -0.40 (-0.41) & 0.04 (-0.05) & & & \\
\ce{O^*} & 0.09 (0.09) & 0.02 (0.02) & -0.05 (-0.04) & 0.07 (0.08) & -0.03 (0.04) & 0.01 & -0.02 \\
\ce{OH^*} & 0.37 (0.35) & 0.05 (0.05) & -0.11 (-0.10) & 0.30 (0.31) & 0.26 (0.3) & 0.02 & 0.26 \\
\ce{OOH^*} & 0.44 (0.46) & 0.08 (0.06) & -0.20 (-0.10) & 0.32 (0.43) & 0.22 (0.38) & 0.03 & 0.23 \\
\hline
\end{tabular}
}
\end{table}

\section{Trends and literature validation}

In addition to comparing to our validation set, we also bench marked the accuracy and performance of our total energy model against the adsorption energies and scaling relationships in literature. By doing so, we hope to demonstrate the utility of our model beyond just our own dataset. All subsequent predictions of literature data were obtained with the \textit{S2EF-total} task using the GemNet-OC OC20+OC22 model with the \textit{Full-ML} scheme in Equation 5. $E_{ad}$ for \ce{O^*}, \ce{H^*}, and \ce{OH^*} was calculated with the following reference schemes:

\begin{equation}
    E_{O^*} = E_{sys} - (E_{slab} + E_{\ce{H2O_{(g)}}} - E_{\ce{H2_{(g)}}})
\end{equation}
\begin{equation}
    E_{H^*} = E_{sys} - (E_{slab} + \frac{1}{2}E_{\ce{H2_{(g)}}})
\end{equation}
\begin{equation}
    E_{OH^*} = E_{sys} - (E_{slab} + E_{\ce{H2O_{(g)}}} - \frac{1}{2}E_{\ce{H2_{(g)}}})
\end{equation}





\begin{figure}
    \centering
    \includegraphics[width=1\textwidth]{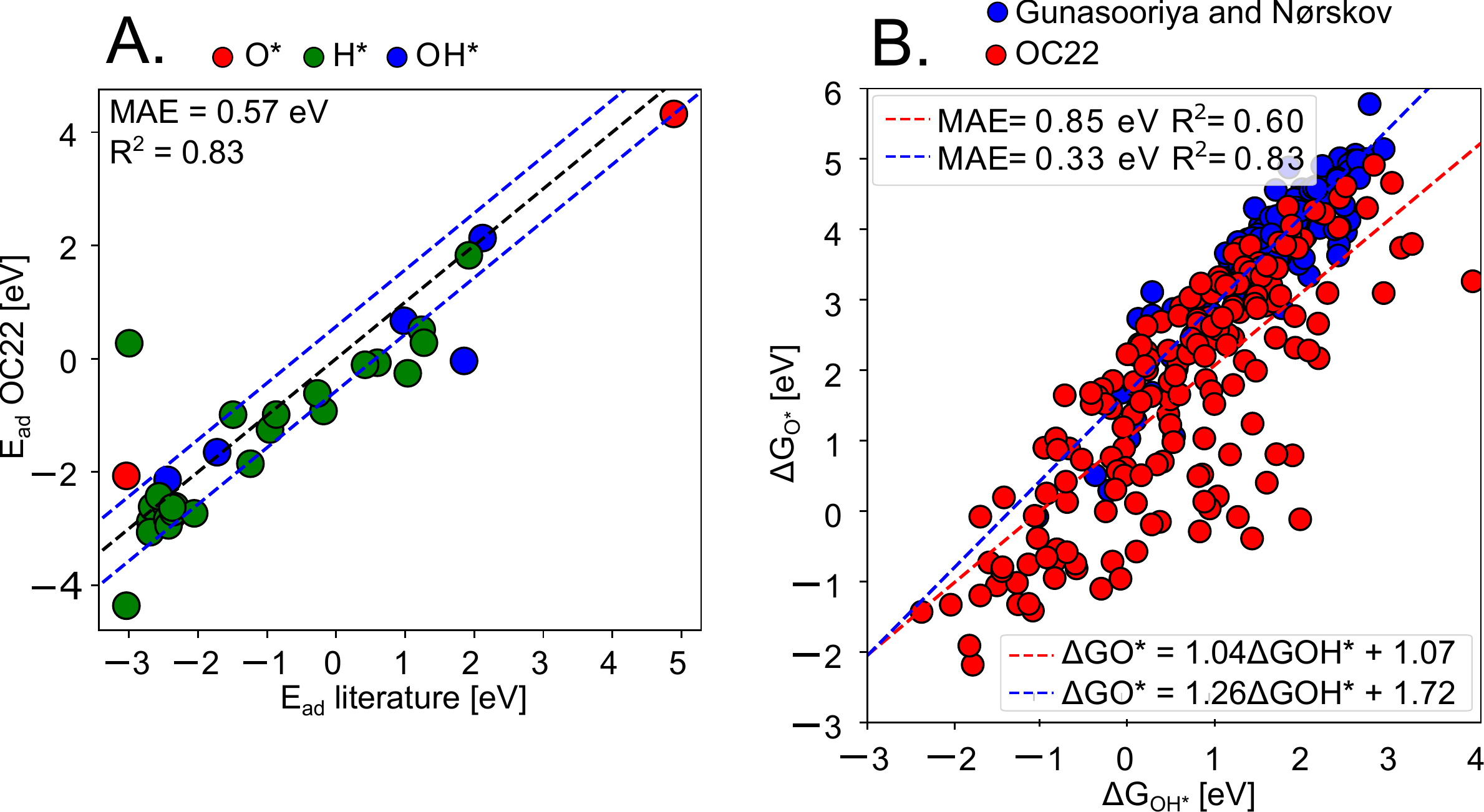}
    \caption{(A) Comparison of OC22 predicted (y-axis) and literature (x-axis) values for the adsorption energies of \ce{O^*}, \ce{H^*}, and \ce{OH^*} across different OOD metal oxide compounds for perovskites structures (see SI for a comparison of perovskite structures)\cite{dickens2019electronic}. A parity line (black-dashed) is provided for reference as well as a line above and below to indicate the mean absolute error (blue-dashed). Literature values were computed using the rPBE functional. (B) A comparison of $\Delta G_{O*}$ (y-axis) and $\Delta G_{OH*}$ (x-axis) with predicted (red) and literature\cite{gunasooriya2020analysis} (blue) data points shown along with their corresponding linear fits (dashed lines). All predictions were performed using the GemNet-OC OC20+OC22 model.}
    \label{fig:lit_compare_si}
\end{figure}

We compared our predicted adsorption energies across different publications, with different functionals, Hubbard U corrections, magnetic configurations, chemical space and materials. We ensure that no combinations of adsorbate+slab are evaluated that exists within the training set. Predictions of $\hat{E}_{sys}$ and $\hat{E}_{slab}$ were calculated using relaxed clean slabs available in the literature as the initial slab configuration in both cases. Figure~\ref{fig:lit_compare_si}(A) plots our predicted values for the adsorption energies of O*, H*, and OH* against literature values for perovskite surfaces\cite{dickens2019electronic}. The datapoints mostly fall within an MAE of 0.57 (blue dashed lines) while maintaining a strong 83\% linear correlation. A similar comparison is made with literature data obtained for rutile structures in Figure 9(A).

We also explored scaling relationships between all combinations of the 9 adsorbates investigated. Scaling relationships were obtained with $\sim$ 180 datapoints with the same $\sim$ 180 clean slabs and adsorbate positions for all adsorbates. The facet, termination, and adsorption site of each of our $\sim$ 180 clean slabs were randomly chosen. The materials used to construct the slabs were given 1/3 splits between materials with elemental combinations out of domain, 60 materials with elemental combinations (but not the crystal structure) in domain, and 60 materials in domain. 

It is known from~\citet{gunasooriya2020analysis} that $\Delta G_{\ce{OH^*}}$ vs $\Delta G_{\ce{OOH^*}}$ and $\Delta G_{\ce{OH^*}}$ vs $\Delta G_{\ce{O^*}}$ have strong linear correlations. We calculated these quantities using $\Delta G_{corr}$ from Table~\ref{tab:gibbs_correction}, for our random set of slabs and plotted the scaling relationships for $\Delta G_{\ce{OH^*}}$ vs $\Delta G_{\ce{O^*}}$ (see Figure 9(B) for $\Delta G_{\ce{OH^*}}$ vs $\Delta G_{\ce{OOH^*}}$) in Figure~\ref{fig:lit_compare_si}(B) alongside the relationship and datapoints obtained from~\citet{gunasooriya2020analysis} for comparison. The slopes of our scaling relationship are consistent within 0.15 eV, however our intercept underestimates the literature value by 0.7 eV.

The slope, intercept and $R^2$ for each scaling relationship of the adsorption energy is also presented in Table~\ref{tab:scaling_relationships}. We obtain $R^2>0.6$ for the scaling relationships of monatomic adsorbates with the exception of $E_{\ce{H^*}}$ vs $E_{\ce{C^*}}$ which had an $R^2$ of 0.55. We also observe moderate linear scaling with $E_{\ce{O^*}}$ vs $E_{\ce{OH^*}}$, $E_{\ce{H^*}}$ vs $E_{\ce{OH^*}}$, and $E_{\ce{OH^*}}$ vs $E_{\ce{OOH^*}}$. All other scaling relationships showed weak or no linear correlation.

\begin{table}
\caption{\label{tab:scaling_relationships} Scaling relationships of the \gls{OC22} adsorbates as evaluated by our best performing total energy model - GemNet-OC \gls{OC20}+\gls{OC22}.}
\resizebox{1\columnwidth}{!}{%
\begin{tabular}{ccccc|ccccc}
\toprule
Adsorbate x & Adsorbate y & Slope & Intercept & $R^2$ & Adsorbate x & Adsorbate y & Slope & Intercept & $R^2$ \\
\hline
\ce{O$^*$} & \ce{H$^*$} & 0.43 & 0.02 & 0.76 & \ce{N$^*$} & \ce{CO$^*$} & 0.25 & 1.68 & 0.26 \\
\ce{O$^*$} & \ce{N$^*$} & 0.78 & 1.17 & 0.88 & \ce{N$^*$} & \ce{H2O$^*$} & -0.04 & 0.09 & 0.01 \\
\ce{O$^*$} & \ce{C$^*$} & 0.59 & -0.15 & 0.62 & \ce{N$^*$} & \ce{OOH$^*$} & 0.54 & -0.26 & 0.44 \\
\ce{O$^*$} & \ce{HO$^*$} & 0.57 & -0.39 & 0.63 & \ce{C$^*$} & \ce{HO$^*$} & 0.65 & -1.06 & 0.46 \\
\ce{O$^*$} & \ce{O2$^*$} & 0.27 & 1.15 & 0.28 & \ce{C$^*$} & \ce{O2$^*$} & 0.38 & 1.02 & 0.28 \\
\ce{O$^*$} & \ce{CO$^*$} & 0.20 & 1.97 & 0.26 & \ce{C$^*$} & \ce{CO$^*$} & 0.25 & 1.78 & 0.21 \\
\ce{O$^*$} & \ce{H2O$^*$} & -0.03 & 0.05 & 0.01 & \ce{C$^*$} & \ce{H2O$^*$} & -0.02 & 0.12 & 0.00 \\
\ce{O$^*$} & \ce{OOH$^*$} & 0.46 & 0.57 & 0.49 & \ce{C$^*$} & \ce{OOH$^*$} & 0.56 & 0.07 & 0.38 \\
\ce{H$^*$} & \ce{N$^*$} & 1.43 & 0.47 & 0.73 & \ce{HO$^*$} & \ce{O2$^*$} & 0.52 & 1.50 & 0.44 \\
\ce{H$^*$} & \ce{C$^*$} & 1.12 & -0.62 & 0.55 & \ce{HO$^*$} & \ce{CO$^*$} & 0.27 & 1.92 & 0.26 \\
\ce{H$^*$} & \ce{HO$^*$} & 1.17 & -0.67 & 0.66 & \ce{HO$^*$} & \ce{H2O$^*$} & 0.02 & 0.24 & 0.00 \\
\ce{H$^*$} & \ce{O2$^*$} & 0.47 & 0.86 & 0.20 & \ce{HO$^*$} & \ce{OOH$^*$} & 0.73 & 0.63 & 0.62 \\
\ce{H$^*$} & \ce{CO$^*$} & 0.41 & 1.84 & 0.19 & \ce{O2$^*$} & \ce{CO$^*$} & 0.12 & 0.93 & 0.07 \\
\ce{H$^*$} & \ce{H2O$^*$} & -0.10 & -0.01 & 0.02 & \ce{O2$^*$} & \ce{H2O$^*$} & 0.26 & 0.18 & 0.16 \\
\ce{H$^*$} & \ce{OOH$^*$} & 0.93 & 0.28 & 0.48 & \ce{O2$^*$} & \ce{OOH$^*$} & 0.52 & -1.55 & 0.42 \\
\ce{N$^*$} & \ce{C$^*$} & 0.83 & -0.90 & 0.85 & \ce{CO$^*$} & \ce{H2O$^*$} & 0.20 & -0.25 & 0.10 \\
\ce{N$^*$} & \ce{HO$^*$} & 0.64 & -1.43 & 0.56 & \ce{CO$^*$} & \ce{OOH$^*$} & 0.58 & -2.80 & 0.40 \\
\ce{N$^*$} & \ce{O2$^*$} & 0.33 & 0.70 & 0.27 & \ce{H2O$^*$} & \ce{OOH$^*$} & 0.25 & -1.58 & 0.09 \\
\hline
\end{tabular}
}
\end{table}

\clearpage
\section{Changelog}

This section tracks the changes to this document since the original release.

\noindent \textbf{v1}. Intial version.

\noindent \textbf{v2}.
\begin{itemize}[noitemsep,topsep=0pt]
    \item Update dataset counts to be consistent with released LMDBs.
    \item Minor text and figure updates for clarity and consistency.
\end{itemize}

\noindent \textbf{v3}. Published ACS Catalysis
\begin{itemize}[noitemsep,topsep=0pt]
    \item For a small subset of systems from OC22 validation adsorption energies were calculated using the conventional DFT procedure.
    \item Adsorption energy predictions from total energy models were benchmarked using the small DFT validation set.
    \item Evaluated the performance of total energy models on literature adsorption energies.
    \item Calculated correction energies for $\Delta G_{O*}$, $\Delta G_{OH*}$, and $\Delta G_{OOH*}$.
    \item Determined scaling relationships between different adsorbates with predicted adsorption energies.
    \item Validated predicted scaling relationships of $\Delta G_{ad}$ with literature data using the correction energies.
    \item Extended our discussion of long-range, magnetic, and charge effects in oxide slabs.
    \item Added additional discussion to address solvation effects and experimental outlook.
\end{itemize}

\end{document}